\documentclass{aastex}

\newcommand{\ms}{$M_{\odot}$}
\newcommand{\msb}{$M_{\odot}$~}
\newcommand{\cd}{$^{12}$C}
\newcommand{\cdb}{$^{12}$C~}
\newcommand{\ct}{$^{13}$C}
\newcommand{\ctb}{$^{13}$C~}

\shorttitle{The $^{85}$Kr branching}
\shortauthors{Abia et al. }

\begin{document}

\title{The $^{85}$Kr $s$-process Branching and the Mass of Carbon Stars}

\author{C. Abia\altaffilmark{1}}
\affil{Dept. F\'\i sica Te\'orica y del Cosmos, Universidad de Granada,
    E-18071 Granada, Spain.}
\email{cabia@ugr.es}

\author{M. Busso\altaffilmark{2}}
\affil{Osservatorio Astronomico di Torino, 10025 Pino Torinese, Italy}
\email{busso@to.astro.it}

\author{R. Gallino\altaffilmark{3}}
\affil{Dipartimento di Fisica Generale, Universit\`a di Torino, Via P. Giuria 1, 10125 Torino, Italy}
\email{gallino@ph.unito.it}

\author{I. Dom\'\i nguez\altaffilmark{4}}
\affil{Dept. F\'\i sica Te\'orica y del Cosmos, Universidad de Granada,
    E-18071 Granada, Spain}
\email{inma@ugr.es}

\author{O. Straniero\altaffilmark{5}}
\affil{Osservatorio Astronomico di Collurania, I-64100 Teramo, Italy}
\email{straniero@astrte.te.astro.it}

\and

\author{J. Isern\altaffilmark{6}} \affil{Institut d'Estudis Espacials
de Catalunya - CSIC, Barcelona, Spain} \email{isern@ieec.fcr.es}

\begin{abstract}

We present new spectroscopic observations for a sample of C(N)-type red giants.
These objects  belong to the class of Asymptotic Giant Branch
stars, experiencing thermal instabilities in the
He-burning shell (thermal pulses). Mixing episodes called {\it
third dredge-up} enrich the photosphere with newly
synthesized \cdb in the He-rich zone, 
and this is the source of the high
observed ratio between carbon and oxygen (C/O $\ge$ 1 by number).
Our spectroscopic abundance estimates confirm that, in agreement with the 
general understanding of the late evolutionary stages of low and
intermediate mass stars, carbon enrichment is accompanied by the
appearance of $s$-process elements in the photosphere. We discuss
the details of the observations and of the derived abundances, focusing in particular 
on rubidium, a neutron-density sensitive element, and on the $s$-elements
Sr, Y and Zr belonging to the first $s$-peak. The critical reaction branching at
$^{85}$Kr, which determines the relative enrichment of
the studied species, is discussed. Subsequently, we compare
our data with recent models for $s$-processing in Thermally
Pulsing Asymptotic Giant Branch stars, at metallicities relevant
for our sample. 
A remarkable agreement between model predictions
and observations is found. Thanks to the different neutron density
prevailing in low and intermediate mass stars,
comparison with the models allows us to conclude that most
C(N) stars are of low mass ($M$ $\la$ 3 \ms). We also
analyze the \cd/\ctb ratios measured, showing that most of them cannot be
explained by canonical stellar models. We discuss how this fact
would require the operation of an $ad$ $hoc$ additional mixing, currently called Cool
Bottom Process, operating only in low mass stars during the first ascent of the red giant branch
and, perhaps, also during the asymptotic giant branch.

\end{abstract}


\section{Introduction}

The production of elements heavier than iron is normally
ascribed to neutron captures. At the low neutron density
prevailing in Asymptotic Giant Branch (AGB) stars, this nucleosynthesis process is referred
to as the {\it s-process}, where the neutron-capture path proceeds
along the valley of $\beta$ stability and most
unstable isotopes encountered along the $s$-path
preferentially $\beta$-decay instead of capturing a further
neutron.  However, in a number of cases, competition between
neutron capture and beta decay leads to a {\it branching} in the
synthesis path, which is determined by the neutron density and
the neutron capture cross section of the unstable nucleus and its
$\beta$-decay half-life (see e.g. K\"appeler, Beer, \& Wisshak
1989). 

The discovery of the unstable element technetium in the
spectra of AGB stars by
\citet{mer52} clearly showed that the $s$-process actually occurs
in these red giants of low to intermediate mass, $1.5\la M/M_\odot\la 8$. The
particular structure of AGB stars (two alternate burning
shells, one of H and an inner one of He, surrounding an inert
degenerate CO core) fulfills the necessary physical conditions
for the $s$-process nucleosynthesis to occur. It is in the He-rich zone 
between the two burning shells (hereafter He intershell)
that $s$-elements are produced by slow neutron captures on seed
nuclei. Periodically the
He intershell  is swept 
by a convective instability induced by a He-burning runaway (thermal
pulse, TP) where $^{12}$C is synthesized. 
 After each thermal instability, newly synthesized $^{12}$C and
$s$-elements are mixed with the envelope by a
mixing episode called {\it third dredge-up} (TDU) \citep{ibe83}.
The repeated action of TPs and TDU episodes during the AGB phase
makes the star to eventually become carbon-rich, i.e. showing a ratio C/O $>1$ in
the envelope. Abundance studies of C-stars (N and SC types) in
the Galaxy have indeed confirmed this figure \citep{uts85, abi98}. 

One debated subject concerning the $s$-process
nucleosynthesis in AGB stars is the knowledge of the neutron
source. Two competing sources have been envisaged: the
$^{13}$C($\alpha,n)^{16}$O and the $^{22}$Ne$(\alpha,n)^{25}$Mg
reactions.  During the last decade a number of
studies in AGB stars of different spectral types have shown that
the $s$-element abundance pattern can only be
understood if the $^{13}$C$(\alpha,n)^{16}$O reaction provides
the bulk of the neutron flux at low neutron densities ($N_n \la
10^7$ cm$^{-3}$) and relatively low temperatures 
($T$ $\sim$ 1 $\times$ $10^8$~K) (cf. Wallerstein et al. 1997; Busso, Gallino, \& Wasserburg
1999, and references therein). 
In AGB stars, this reaction occurs in radiative conditions during the interpulse
phase, in the top layers of the He intershell
\citep{str95, str97, her97, gal98, her00}. In fact, the convective instability powered by a TP 
spreads over the whole He intershell the $^{12}$C produced by the 3$\alpha$ reactions. 
When TDU takes place, a few amount of protons 
may diffuse from the H-rich envelope down to the He intershell. Then, during the interpulse, 
the region where these protons are stored heat up so that 
$^{13}$C is rapidly formed, via proton captures on $^{12}$C, and later on,
when the temperature approaches $10^8$ K, 
the $^{13}$C$(\alpha,n)^{16}$O neutron source is activated \citep{bus99}.

On the other hand,  the $^{22}$Ne$(\alpha,n)^{25}$Mg neutron source requires
higher temperatures ($T \ga$ $3.5$ $\times$~10$^8$~K), as those 
currently found in the TP of intermediate
mass models (i.e. $M$ $>$ 4 $M_\odot$) (see e.g. Straniero et al. 2001). 
The maximum temperature achieved in low mass AGB stars (LMS) 
($M~\la~3$~$M_\odot$)  instead does not exceed $T$ $\sim$ 3
$\times$ $10^{8}$~K (Iben \& Truran 1978; Straniero et al. 1997). Therefore,
 in LMS the $^{22}$Ne neutron source is only marginally
activated, and the $^{13}$C source is the main supplier of
neutrons. Conversely, in AGB stars of $M\ga 4$ $M_\odot$, both 
the $^{22}$Ne and the $^{13}$C neutron sources are efficiently 
activated, during
the thermal pulse and during the interpulse, respectively.

The neutron burst by the $^{22}$Ne neutron source occurs typically at high peak
neutron densities, $N_n \ga 10^{10}$ cm$^{-3}$ (see Vaglio et al. 1999; Busso, 
Gallino, \& Wasserburg 1999). This strongly favors the production of
the neutron-rich nuclides affected by the branching in
the $s$-path, like $^{86}$Kr, $^{87}$Rb and $^{96}$Zr. Therefore,
a different $s$-element pattern is expected depending on
whether one or the other neutron source is more active, thus allowing
us to estimate the neutron density at the $s$-process site and consequently the
initial AGB mass. The typical mass of the C-stars may be inferred by means of 
their luminosity distribution.
In the LMC the C-stars luminosity function is peaked at $M_{\rm bol}$ ${\sim}$ $-$4.5
(see e.g. Hughes \& Wood 1990). A higher value ($M_{\rm bol}$ ${\sim}$ $-$5) 
is found in the Milky Way. These luminosities 
are typical of low mass AGB stars. Very few C-stars have been found at $M_{\rm bol}$ 
$<$ $-6$ (van Loon et al. 1998, 1999; Wallerstein \& Knapp 1998; Trams et
al. 1999), the mass of C-stars in the Magellanic Clouds being estimated to be lower than 
2.7~\msb (Frogel, Mould, \& Blanco 1990). Such an occurrence seems to exclude that 
stars with $M\ga 4~M_\odot$ could become C-stars.
However, it may be possible that the lack of bright C-stars is due to an observational 
selection effect: the brightest AGB stars could be obscured by an optically thick wind before they reach
the C/O $>$ 1 phase. 
On theoretical grounds, the question of the C-star mass has been a debated subject, 
because of the difficulty of finding TDU in LMS (Iben 1981). This problem has by 
now been overcome, since the TDU occurrence has been found in AGB stars with initial 
mass as low as 1.5 $M_\odot$ \citep{lat89, str97, her00}.

In addition, AGB stars of intermediate mass develop a large core 
mass ($M_{\rm H}>0.9$ $M_\odot$), and 
the temperature attained at the base of the convective envelope during the interpulse 
period becomes larger than $70-80\times 10^6$ K, so that $^{12}$C is converted into 
$^{14}$N. This is the so-called Hot Bottom Burning (HBB, Sugimoto 1971; Iben 1975; 
cf. Iben \& Renzini 1983) which is currently found in the most recent computations 
of intermediate mass AGB models (e.g. Bl\"ocker \& Sch\"onberner 1991; 
Boothroyd, Sackmann, \& Ahern 1993; Lattanzio et al. 1996;  Forestini \& Charbonnel 
1997; Straniero et al. 2001). Such an occurrence could explain the lack of bright 
C-stars. However, the HBB efficiency also depends 
on the envelope mass. When, as a consequence of the mass loss, the mass of
the envelope is 
reduced to $\sim$1.5 $M_{\odot}$, HBB eventually ceases, while TDU
continues to mix carbon from the He intershell with the surface (Boothroyd
\& Sackmann 1992; Frost et al. 1998). This could lead to the formation of a bright, 
possibly obscured, 
C-star.  Note that during the HBB phase the \cd/\ctb ratio would remain very low 
and close to its equilibrium value of 3.5, and eventually rise up to $5-10$ when C/O $\sim$
1 is achieved in the envelope (Lattanzio \& Forestini 1998).

In this study abundances of Rb, Sr, Y and Zr in a sample of Galactic C-stars 
are simultaneously derived for the first time. By comparing these findings
with those predicted by $s$-process nucleosynthesis models at the $^{85}$Kr-branching
in AGBs of different masses, we infer about the stellar mass of C-stars 
and the neutron source playing the main role in the $s$-process. In $\S~2$ we describe
the properties of the $^{85}$Kr-branching. $\S~3$ presents the observations and 
analysis. In $\S~4$ the abundances derived in our sample of C-stars are discussed and compared with 
the $s$-elements abundance pattern found in others AGB-stars. Finally, in $\S~5$ and $\S~6$
we discuss the observational results in the framework of the current $s$-process nucleosynthesis
models in LMS and IMS. Our conclusions are summarized in $\S~7$.

\section{The reaction branching at $^{85}$Kr and the neutron density}

The peculiarities of the neutron flow near the {\it magic} neutron
number $N$ = 50, and through the unstable $^{85}$Kr in particular,
offer the possibility of making an observational estimate of the neutron
density. Neutron capture on $^{84}$Kr leads either to the 10.7 yr
ground state of $^{85}$Kr or to the 4.5 h isomeric state at 350
keV, which $\beta$-decays roughly by 80$\%$ to $^{85}$Rb and by
20$\%$ to the $^{85}$Kr ground state through a
$\gamma$-transition. The production rate of $^{85}$Kr$^{iso}$ is
quite important, amounting to about 50$\%$ of the total neutron
capture cross section on $^{84}$Kr. The half-life of the ground
state of $^{85}$Kr is long enough for the $s$-process flow 
to proceed further to $^{86}$Kr and then to
$^{87}$Rb, in competition with $\beta$-decay to $^{85}$Rb,
depending on the neutron density. When the neutron density is
higher than a few 10$^8$ cm$^{-3}$, the neutron channel is open
and the $s$-process path feeds the neutron magic nuclei $^{86}$Kr
and $^{87}$Rb. Since $^{85}$Rb has a neutron capture cross section larger 
by a factor $\sim 10$ than  $^{87}$Rb, the total abundance
of Rb as well as its isotopic mix is very sensitive to the
neutron density. Indeed, the total Rb abundance can differ by one order of
magnitude depending on whether the low or the high neutron density routing is
at work (see Beer \& Macklin 1989 for a detailed description of
this branching). Therefore, the relative abundance of Rb to other
elements in this region of the $s$-process path, such as Sr, Y
and Zr, can be used to estimate the average neutron density of
the $s$-process and, as a consequence, to infer the mass of the
star. In fact, in the last 15 years this kind of study for Rb has
been successfully performed in field stars \citep{tom99} and in
AGB stars of different spectral types: Ba stars \citep{tom83,
mal88}, MS and S giants \citep{lam95}, yellow symbiotics
\citep{smi97} and some $s$-process-rich stars in $\omega$ Cen
\citep{van94, smi00}. These abundance studies have added further
evidence favoring the $^{13}$C$(\alpha,n)^{16}$O reaction as the
neutron source. 
The observed sources in the above studies
often belong to binary systems in which the companion is now a
white dwarf \citep{jor93}. Thus, they owe the $s$-element
overabundances to mass transfer from an ancient AGB star (the
progenitor of the white dwarf companion) and not to an ongoing
$s$-process. 
As to C(N) stars, the only spectroscopic estimates of $s$-process elements  
available so far are those by Utsumi (1985), which are however based on low-dispersion
photografic plates, nor Rb was measured at the time. With the improved set of 
spectroscopic data presented here we are able to determine the Rb abundance 
and to provide a much better estimate of Sr, Y, Zr, thus allowing us to
examine the general problem of the initial mass also for the class of the
C(N) stars.

\section{Observations and analysis}

The spectra of the C-stars analyzed in this study were taken at
the Calar Alto Observatory using the 2.2 m telescope, and at the
La Palma Observatory with the 4.2 m William Herschel telescope
(WHT). They form part of a more extended study concerning the
chemical composition of Galactic C-stars: see \citet{abi98},
hereafter Paper~I, and \citet{abi00}, hereafter Paper II. The echelle
spectra obtained have a continuous coverage from $\lambda\sim$
4000 to 9000 {\AA} and a resolution
$\lambda/\Delta\lambda$ = 40000 for the Calar Alto spectra, or
60000 for those taken at WHT. The spectra are of good quality
with $S/N$ ratios from $\sim 50$ in the blue part of the spectrum,
up to more than 100 in the red part. The spectra were reduced
following standard techniques using the IRAF software package.
To remove telluric absorptions, the reduced and calibrated spectra of the target stars were
obtained by subtracting the spectrum of a bright, hot and rapidly rotating
star.

The stellar parameters of the stars studied were derived in a
similar way as in Paper II. In short: effective temperatures were
mainly deduced from the calibration of the observed bolometric
flux to the infrared flux in the $L'-$band ($3.7 \mu$m) versus
$T_{\rm{eff}}$ (stars marked with an asterisk in Table 1),
or from the $(J-L')_o$ vs. $T_{\rm{eff}}$ calibration by \citet{ohn96}. Infrared
photometry was taken from \citet{fou92} and \citet{nog81}. When photometry was not
available we adopted the $T_{\rm{eff}}$ values derived from angular 
diameter measurements \citep{dyc96}. Otherwise, we adopted the
value given in \citet{lam86} for some stars in common with this
work. We have used the grid of model atmospheres for C-stars
computed by the Uppsala group (see Eriksson et al. 1984, for details).
These models cover the $T_{\rm{eff}}$ = $2500-3100$ K, C/O $=$ $1.0-1.35$
ranges and all have the same gravity, log~$g$~=~0.0. Considering
the $T_{\rm{eff}}$ of the program stars and a typical luminosity
10$^4$ $L_\odot$, this is a common value for AGB stars.
Nevertheless, our gravity might be in error by as much as $\pm~1$~dex 
due mainly to the uncertain estimate of the luminosity
($M_{\rm{bol}}$). The CNO abundances in the model atmosphere of
a given star were taken from the literature when available
\citep{lam86}, or estimated from spectral synthesis in the
wavelength ranges analyzed here. The absolute abundances of CNO
that can be estimated in that way are rather uncertain; however, the C/O ratio
that mainly determines the global shape of the spectrum, is
derived with good accuracy ($\sim \pm 0.05$). Whenever possible,
$^{12}$C/$^{13}$C ratios were obtained from our echelle spectra
by synthetic fits to $^{13}$CN features in the
$7990-8040$ {\AA} range following the same procedure as in
\citet{abi96}. Otherwise, they were taken from the literature
\citep{lam86, ohn96}. Some of the $^{12}$C/$^{13}$C ratios derived
here differ from those in \citet{abi97} because of the different
$T_{\rm{eff}}$ values and an updated CN and C$_2$ line list in the
$7990-8040$ {\AA} range used here. Finally, a
typical microturbulence parameter of $\xi\sim 2.2$ km s$^{-1}$ was
adopted in the analysis of all sample stars \citep{lam86}. Table 1
shows the atmosphere parameters for the program stars.

The metallicity [M/H]\footnote{We adopt here the usual notation
[X] $\equiv$ log (X)$_{program-star}$ $-$ log (X)$_{comparison-star}$ for
the abundance of any element X.} of our sample sources was derived
from selected metallic lines in the stellar spectra (see Table 3).
Most of these lines are located in the wavelength ranges
$4450-4650$ {\AA} and  $4700-4950$ {\AA}. In these wavelength windows, molecular 
absorptions are weak in C-stars and metallic lines can be identified more easily. The
metallic lines were selected according to the same careful criteria described 
in Paper I. Most of the $gf$-values were taken from the literature or from
the VALD \citep{pis95} and Kurucz's CD-ROM No.~23 data bases. In some cases, we
derived solar $gf$-values using the \citet{how74} atmosphere model 
for the Sun. The abundances of metals were derived by the usual method of 
equivalent width measurements and curves of growth calculated in LTE. Upper limits to
the equivalent widths (see Table 3) were not considered in deriving abundances. The [M/H]
value for each star in Table 2 is the mean value obtained from the selected 
metallic lines.

The abundances of the low-mass $s$-elements around the
$^{85}$Kr-branching (Rb, Sr, Y and Zr) were derived from
spectroscopic features selected following the same criteria as
those for metallic lines. Despite the fact that our C-stars show
many heavy-element absorption lines, only a few of them were
considered useful for abundance analysis. The reason of this
is the severe blending with CN and C$_2$ lines, which is evident
even at the high spectral resolution used here. The lines selected 
are shown in Table 3. Spectroscopic parameters 
were taken from the literaure (see Table 3) since due to their weakness 
in the solar spectrum, the derivation of solar $gf$-values is very 
uncertain. Nevertheless, our $gf$-values are in very good agreement with
the solar $gf$-values list derived by Th\'evenin (1989, 1990). For a
sample of 30 lines in common we found a mean difference of
0.12$\pm 0.20$ dex. In a few cases, however, there is a significant difference 
which, in our opinion, clearly shows that the derivation of solar $gf$-values from
very weak lines is not always safe. For instance, our main Y abundance indicator 
is the Y I line at 4819 {\AA}. This feature is quoted with a 3 m{\AA} absorption in the 
Solar Spectrum Atlas by \citet{moo66}. However, Hannaford et al. (1982) note that the uncertainty
in the wavelength of this line could be as high as 25 m{\AA} and very
probably, Y being not the main contributor to this absorption in the Sun\footnote{Indeed,
the identification of this line as Y in the Solar Spectrum Atlas of \citet{moo66} is
quoted with a question mark.}. In fact, using a solar $gf$-value for this line 
($\sim 1$ dex higher than the value in Table 3) in the analysis of WZ Cas 
(our reference star, see below), we obtain a very low Y abundance ([Y/H]$\sim -1$) which 
is unrealistic given the near solar metallicity of this star and its level of heavy element 
enhancement (see Paper II). The same figure is found with two additional discrepant Ti I lines with the
Th\'evenin's list (at 4812 and 4821 {\AA}): unrealistic Ti abundances are obtained in WZ Cas if solar 
$gf$-values are used (the absorptions in the Sun are 8.5 and 3 m{\AA}, respectively). Finally, our  
$gf$-value of the Fe I line at 5848 {\AA} also differ significantly with 
that by Thevenin. However, the iron abundance derived from this line in the only star
where it is identified (Z Psc), is in agreement within $\pm 0.2$ dex with that derived from 
other Fe lines in this star. 
 
Abundances of s-elements were also derived from curves of growth in LTE except
for those lines presenting clear blends, namely: the
7800 {\AA} Rb I, 7070 {\AA} Sr I and 4815.05, 4815.65 {\AA} Zr I lines, for which we 
used spectral synthesis (see Table 3). CN and C$_2$ lines in these wavelength ranges were
kindly provided by B. Plez and P. de Laverny according to the
spectroscopic predictions by \citet{del98}. Final individual
abundances were obtained as the mean of the abundances derived
from each line. Unfortunately, given the important error in the
measurement of the equivalent widths and the small number of lines
used, it was impossible to check the atmospheric parameters
($T_{\rm{eff}}$, log $g$ and $\xi$) using the requirement that
individual abundances derived from lines of different intensity,
ionization state and excitation energy, have to be nearly equal.

In order to reduce systematic errors due to wrong {\it gf}-values and
departures from LTE, certainly present in late-type giants
\citep{tom83}, we performed a differential analysis line-by-line using WZ Cas 
(C9,2J; SC7/10) as the comparison star. This star has a similar $T_{\rm{eff}}$ and 
surface gravity as the stars studied here and its spectrum is also dominated by CN and C$_2$
absorptions. However, because the C/O ratio in its atmosphere
is very close to unity (C/O $\approx$ 1.005) the spectrum is less
crowded with molecular absorptions, and thus the atomic line
identification is easier. This permits a more accurate abundance
analysis. Abundance studies of this star reveal that its
metallicity is almost solar, [Fe/H] $\approx$ 0.0 (Lambert et al.
1986; Paper II), with no or very small $s$-element overabundances:
the mean heavy element enhancement is [h/Fe]$= 0.02 \pm 0.20$ , where `h' includes Sr, Y, 
Zr, Nb, Ba, La, Ce, and Pr (see Paper II). All the abundance ratios shown in Table 2 are relative to 
WZ Cas\footnote{The [h/Fe] value in WZ Cas differs slightly from that derived in Paper  
II [h/Fe] = $+0.16$, because of the downward revised Sr, Ba and La abundances here.
This new analysis was done with a very high resolution spectrum (R$\sim 180000$,
S/N$>70$) obtained, for another scientific purpose, with the 2.5 m NOT and the  
SOFIN echelle spectrograph at La Palma Observatory. In the re-analysis of WZ Cas we have used 
Sr, Ba and La lines mainly placed in the $4800-4950$ {\AA} spectral range.}.

The derivation of the Rb abundance requires special attention. In
C-stars the line used here is strongly blended with CN, C$_2$
lines and some atomic absorptions. Therefore, the Rb abundance
must be derived by spectral synthesis. We have used an
updated version of the molecular and atomic line list used in
Papers I and II. The new line list includes some additional
$^{12}$C$^{15}$N, $^{13}$C$^{15}$N and $^{13}$C$^{13}$C lines
computed by B. Plez. We have also modified the {\it gf}-value of a weak
Si I line to the blue of the Rb line according to \citet{tom99}.
The hyperfine structure of the Rb I line was considered adopting
a Solar System isotopic ratio $^{85}$Rb/$^{87}$Rb = 2.59
\citep{and89}. Note that the isotopic Rb mixture can certainly be
altered as a function of the neutron density in the stars.
Unfortunately, this isotopic ratio cannot be measured
spectroscopically. Nevertheless, we checked that other isotopic
mixtures will not change significantly our conclusions (see
below). We have adopted the meteoritic Rb abundance, log~(Rb/H)~+~12 = 2.40
$\pm$ 0.05 \citep{and89}. We preferred to adopt the
meteoritic abundance rather than the photospheric one ($2.60 \pm
0.15$) as the reference value because of the weakness of the
7800~{\AA} Rb I line in the Sun ($<1$ m{\AA}) from
which the photospheric abundance is derived. Instrumental and
macroturbulence broadening were included in the Rb synthesis. The
macroturbulence broadening was set by matching the profile of an
almost clean nearby Ni I line at 7788.9 {\AA}.

A striking result immediately appears in the analysis of the
7800 {\AA} Rb I line: the Rb abundances derived  are
unrealistically low considering the near-solar metallicity of
most C-stars analyzed here. In fact, the derived
abundances of Rb are lower than solar by a factor ranging
typically between four and ten. This figure has also been found in
similar studies on AGB stars \citep{ple93, lam95, abi00}. The
reason for this is not well understood, although N-LTE effects due
to overionization might partially explain this finding. In fact,
due to its low ionization energy (4.16 eV), rubidium is mostly
ionized at the depth of the 7800 {\AA} line formation.
\citet{abi99} showed that N-LTE effects in C stars due to
overionization in the formation of the resonance Li I at
6708 {\AA} (another alkali element with an ionization energy
similar to Rb, 5.39 eV) might extend up to +0.6 dex (in the sense
of N-LTE minus LTE abundances). Surprisingly, the same effect is
found when deriving K abundances (ionization energy 4.34 eV) from
a nearby resonance K I line at 7698.70 {\AA}; the K
abundances derived are also unrealistically low: typically
[K/H] $\sim$ $-$1\footnote{Certainly N-LTE effects are important in
the formation of the resonance 7698 {\AA} K I line in
the Sun \citep{tak96}. However, as far as we know, there is no a
similar study in C-rich giants.}. No nuclear process able to
deplete K in stars is known. Although both the Li and K lines are
usually stronger than the Rb I line in C-stars and, in
consequence, they should form higher in the atmosphere where
larger departures from LTE are expected, upward corrections of
the Rb abundance are not excluded until a quantitative analysis
of N-LTE effects in the resonance Rb line is made. However, we
believe that N-LTE effects are not probably the full story. In
our opinion there also seems to be a problem related to the
continuum opacity in the 7800 {\AA} region. Indeed, the
metallicity derived from (almost unblended) Ni~I and Fe~I lines
near the Rb I line is significantly lower (by a few tenths of dex) 
than the one derived from the metallic lines in the blue part of the spectrum. This
suggests that the continuum absorption coefficient of the real
atmosphere in the Rb spectral region is not well reproduced by
our model atmospheres. Tests made by modifying the model atmosphere
parameters in a quantity similar to the expected errors (see
Table 4) do not solve the problem. However, when trying to
artificially simulate this missing opacity by increasing the
H$^-$ opacity (namely, increasing the electronic pressure in the
model atmosphere), the derived metallic and Rb abundances
increase by tenths of dex. Again, we have to wait for improved model
atmospheres of C-stars \citep{ple99} to study this
problem better. Therefore, because of the uncertain Rb abundances, we
computed the [Rb/M] ratio using the metallicity ([M/H]) derived
from the metallic lines close to the Rb line. The [Rb/M] ratios
in Table 2 were computed in that way, also using WZ Cas as the
comparison star.

Unlike Rb, strontium has many 
spectroscopically accessible lines. However, in C-stars most of them are very
strong and/or located in crowded regions of atomic and molecular
absorptions. As the main indicator of the Sr abundance we have
used the Sr~I line at 7070.1 {\AA}. In some of the
stars observed at a very high resolution ($R\sim 60000$), it was
possible to derive Sr abundances from two additional Sr I
lines at 4811 and 4872 {\AA}. In the neighbourhood of these lines, however, the 
continuum position is uncertain. The Sr I line at 7070 {\AA} is
blended with some $^{13}$CN lines of moderate intensity, so
that spectral synthesis is required. We obtain a solar
$gf$-value for this line from the solar spectrum using the
photospheric Sr abundance $2.90 \pm 0.06$ \citep{and89}. No
hyperfine structure was considered for the 7070 {\AA} Sr~I
line, since this absorption is rather weak (log~$W_\lambda/\lambda$~$\la$~$-4.5$) 
even in C-stars with large Sr enhancements. Due to the
blend with $^{13}$CN lines, the abundance of Sr derived from this
feature is sensitive to the $^{12}$C/$^{13}$C ratio adopted.
Unfortunately, given the current uncertainties in the derivation
of CNO abundances and the atmospheric parameters of C-stars, the
typical error in the derivation of the $^{12}$C/$^{13}$C ratio is
still high, $\pm$($6-13$): the higher the $^{12}$C/$^{13}$C ratio,
the higher the error (see Abia \& Isern 1996). This introduces
an extra uncertainty into the derivation of the Sr abundance from
this line, mainly in stars with $^{12}$C/$^{13}$C $\la 30$.
Therefore, the [Sr/M] ratios derived here have to be considered
with some caution\footnote{Nevertheless, in the stars where it was
possible to derive Sr abundances from the lines in the blue, the
dispersion never exceeded $\pm0.3$ dex, which is consistent with the expected total error.}. 
On the other hand, synthetic fits in the Sr spectral range include a clean Ti
I line at 7065.50 {\AA} that was used as an additional
check to the [M/H] ratio derived from metallic lines in the blue
part of the spectrum and to set the macroturbulence broadening.

\subsection{Errors}

Errors involved in the derivation of abundances in C-stars are large.
The sensitivity of the derived abundance ratios to the model
atmospheres was assessed by changing the atmospheric parameters
over a plausible range of $T_{\rm{eff}}$, $\xi$, CNO
abundances, C/O and $^{12}$C/$^{13}$C ratios. Obviously, for a
given element the change in the abundance created by varying each
atmospheric quantity depends on the excitation energy and on the
intensity of the specific line (we have used only neutral lines).
In Table 4 we show the sensitivity of the [X/H] ratios to the
above sources of error. We have also included the error
introduced by an uncertainty of 5$\%$ in the location of the
continuum in the spectrum (if spectral synthesis is used) or due
to errors in the equivalent width measurements (when curves of
growth are used). We do not include the effect of uncertainties in the
gravity since we have only atmosphere models computed with log $g=0.0$.
However, since we have used only neutral lines in the analysis, our abundances
would not be very sensitive to changes in gravity within a variation of $\pm 0.5$ dex in
log $g$. The values reported in Table 4 were computed using
equivalent widths and model atmosphere parameters that are
representative in the program stars, namely: $-$log
$W_\lambda/\lambda = 4.3$ to $5.0$, $T_{\rm{eff}} = 2850$ K, log $g = 0.0$,
$\xi= 2.2$ km s$^{-1}$, C/O = 1.05 and $^{12}$C/$^{13}$C = 40.
From this table it is apparent that the mean sources of error come
from uncertainties in $T_{\rm{eff}}$, microturbulence and
errors in the equivalent width measurement or the
spectral synthesis fit (by eye). Indeed, the microturbulence in variable stars might 
differ with wavelength in the spectrum (due to differences in depth of line formation), and might
change with the pulsation phase of the star. We believe, however, that an uncertainty in 
the microturbulence as large as $\pm 1$ kms$^{-1}$ 
can be discarded because this can be checked from synthetic fits to CN lines in the spectral 
regions studied. For instance, synthetic fits to the strong and crowded CN lines in the 
$7070$ {\AA} region show that such a variation respect to our adopted value 
(2.2 kms$^{-1}$) would either result in CN lines that are too narrow and free of significant 
blending with adjacent lines, or produce extreme saturation effects. In the former case, the
convolution parameters required to fit the spectrum 
would be very large. In the later, the resulting spectrum would be very smooth with an unrealistic 
continuum position. Furthermore, the effect of variations in the microturbulence may also be 
distinguished from that of variations in the C/O ratio since the former deepen and become 
broader and more blended with the neighboring features. We note that Lambert et al. (1986), 
analyzing weak CO and CN lines (log ($W_\lambda/\lambda)\la -4.5$) in a sample of 30 C-stars 
(many of them studied here), obtain a mean value $\xi=2.16\pm 0.20$ kms$^{-1}$. 
On the other hand, note from Table 4 the small sensitivity of the Y and Zr abundances and 
metallicity to uncertainties in the C/O, $^{12}$C/$^{13}$C and CNO/H ratios.
This is because these abundances are derived using curves of growth from 
unblended (with CN or C$_2$ features, as far as our molecular line lists indicate) 
lines.

The quoted accuracy of our $gf$-values range between $5\%-25\%$, which is translated to an
error in the derived abundances between 0.05-0.25 dex depending on the intensity of the line.
However, as all the abundances derived here are set in ratio to WZ Cas, errors in the
$gf$-values are not the relevant ingredient. More important is the random
error in the abundance of elements represented by a few number of lines or just one. In this
case, the quality of the lines in question is the critical point. For instance, we consider the 7800 {\AA} Rb I line
as a good indicator since our synthetic spectra reproduce quite well this spectral region.
The same applies for the Zr I blend at 4815 {\AA} and for the Y I line at 4819 {\AA}, which
appears safely clean of blends in the majority of the spectra. In any case, an estimate of this
random error can be evaluated from the resulting dispersion around the mean abundance value in the
cases where more than three lines are used. This can be done for a few stars (see Table 3) and only 
for Ti and Fe abundances. We found a dispersion in the range 0.1-0.25 dex. On the other hand, systematic errors, 
not included in Table 4, may be present affecting the [X/M] ratios. The leading contributor to a
systematic error is probably our assumption that N-LTE effects
are reduced when the program stars are analyzed with respect to WZ
Cas. Note also that our stars are variable, therefore, the
effective temperature might change with the pulsational phase.
The quoted $\Delta T_{\rm{eff}}$ in Table 4, thus, refers to the
typical expected uncertainty in the derivation of this parameter
according to the different methods used: photometry, infrared flux or angular 
diameter measurements (see above). Focusing on Rb for instance, a Rb isotopic 
mix different than solar would modify the derived Rb abundance. Nevertheless,
variations of the $^{85}$Rb/$^{87}$Rb by a factor of five implies a
maximum variation of only $\pm0.1$ dex in the Rb abundance. On
the other hand, if the solar photospheric Rb abundance is
preferred as a reference, the [Rb/M] ratios in Table~2 have to be
reduced by 0.2 dex. Other sources of error such as the uncertainty in
the model atmospheres (hydrostatic plano-parallel approximation,
velocity stratifications, poly-atomic opacities), dust or
chromospheres are at present impossible to quantify. We estimate
a total typical error (non systematic) of $\pm0.30 $ dex
for [M/H], $\pm0.30$ dex for [Rb/H], $\pm0.35$ dex for [Sr/H],
$\pm0.30$ dex for [Y/H], and $\pm0.40$ dex for [Zr/H]. Adding-up quadratically to  
these figures the random error due to the use of a few number of lines, the total
error in the [X/H] ratios would range between $\pm$0.3-0.45 dex. Errors in the [X/M] and
[X/Y] ratios are, in general, lower since these ratios are less sensitive than the 
[X/H] ratios to the atmosphere parameters: sources of error affecting in
the [X/H] ratios to the same sense cancel out when computing these  
element abundance ratios.

Below, we compare our results with previous abundance analyses and with theoretical
predictions of $s$-process nucleosynthesis in C-stars of different masses.

\section{The abundance ratios and their characteristics}

A detailed chemical analysis of heavy element abundances from Rb
to Ce in a larger sample of Galactic C(N)-stars will be presented
in a forthcoming paper (Abia et al. 2001, in preparation). Here
we restrict the discussion to the low-mass $s$-elements (Rb, Sr, Y
and Zr). Table 2 shows the abundance ratios with respect to the
metallicity derived in our stars referred to WZ Cas. Most 
stars are of near-solar metallicity, except V CrB and IY Hya, which 
can be considered as metal-poor C-stars (see below). 
The stars present moderate light $s$-element (Zr, Y, Sr) enhancements,
namely: $<$[Rb/M]$>$ = 0.27 $\pm$ 0.24, $<$[Sr/M]$>$ = 0.41 $\pm$
0.19, $<$[Y/M]$>$ = 0.59 $\pm$ 0.17 and $<$[Zr/M]$>$ = 0.54 $\pm$
0.22. These overabundances are significantly lower than those
reported by \citet{uts85} in a similar analysis. Excluding Rb (not
studied by Utsumi), our [ls/M] ratios, with `ls' = (Sr, Y, Zr),
are $0.4-0.6$ dex smaller. However, as discussed in Paper II, the abundances
derived by Utsumi (1985) were based on atomic line
identifications made in a previous work (Utsumi 1970), and indeed
most of the heavy-element lines identified in this work appear as
clear blends at the spectral resolution used here (a factor three
higher than Utsumi's one). This necessarily leads to abundance
overestimations. Additionally, Utsumi used in his analysis the 
Allen's (1976) Solar System abundances as reference values, which
are lower by 0.05 (Sr), 0.44 (Y) and 0.1 (Zr) dex than the
reference values used here \citep{and89}. This explains part
of the differences found with respect to Utsumi's analysis.

In any case, the light $s$-element abundance ratios derived here
are lower than previously thought and contrast with the higher
overabundances found in other C-rich $s$-enhanced stars or their
descendants (Ba , CH and SC-stars: see for instance Vanture 1992a,b; 
Smith \& Lambert 1988), and in fact are more similar to the findings of
O-rich S stars (Smith \& Lambert 1990). 
We remind that during the AGB phase it is commonly assumed that the
abundance of $^{12}$C and $s$-elements increases in the envelope
of the star along the spectral sequence
M$\rightarrow$MS$\rightarrow$S$\rightarrow$SC$\rightarrow$C. 
As the TP and TDU episodes occur repeatedly, and since most of our
stars have C/O ratios near unity, they should show abundances similar
to those found in SC-stars\footnote{SC-stars are AGB stars with a
C/O ratio very close to unity within 1$\%$ or less. Furthermore, they 
show spectroscopic characteristics which are used to distinguish them from
the C(N)-stars (see Keenan \& Boeshaar 1980, for a complete definition).}. 
However, comparison with the average low-mass $s$-element
enhancements found in SC-stars (see Paper I) shows a difference of
about $+0.5$ dex, in the sense of SC-stars minus C(N)-stars. We believe that
this difference is due to several reasons.
First, the abundances in Paper I were derived from lower resolution
spectra (a factor $\sim 2$). Second, despite the fact that SC stars show less
intense molecular absorptions (CN \& C$_2$), due to their C/O 
ratio being very close to unity, the abundance analysis in Paper I was
based on atomic lines identified in spectral ranges ($5900-8000$~{\AA}) 
that were much more crowded with molecular absorptions than
those mainly used here ($4750-4950$~{\AA}). These two facts
might have produced systematic abundance overestimates due to
undetected blends. However, the main reason is probably due to differences
in the model atmospheres used. In Paper I atmosphere models for
SC-stars from D. Alexander (private communication) were used. We
checked that for the same atmosphere parameters, the thermal
structure in the higher layers differs significantly from that
obtained in the Uppsala models used here. In fact, the abundances
derived using Alexander's models differ from those derived using
the Uppsala models by 0.4 to 0.7 dex!, in the sense of Alexander
minus Uppsala abundances: the lower the excitation energy of the
line, the higher the abundance difference is. This affects
the $s$-elements abundance estimates more than the metallic abundances, since the
latter are derived from higher excitation energy lines. If this
effect were considered, the [X/M] ratios in our stars would have to be
increased (or decreased in SC-stars) systematically by 0.3 to 0.5 dex 
depending on the individual element. In that case SC- and
C(N)-stars $s$-element overabundances can be estimated at similar
levels within the error bars. This fact is a clear example of the
large uncertainties involved in the chemical analysis of AGB
stars. 

As to the class of MS/S stars, which are almost as close to solar
metallicity as the present sample of C-stars, the average
$<$[ls/Fe]$>$ is $+0.7$, with a typical uncertainty of 0.25 dex
(Busso et al. 2001). Within the respective uncertainties they
have therefore enhancements very similar to those of C(N) stars
in the present paper. With respect to the warmer BaII giants in the disk,
which show $<$[Fe/H]$>$ = $-0.3$, the subsample with
[Fe/H]~$>$~$-0.3$, amounting to 8 stars in the selection made by
Busso et al. (2001), gives $<$[ls/Fe]$>$ = +0.85. Actually, the Ba stars
are considered to be extrinsic AGBs, i.e. binary systems in which
the present giant was enriched in $s$-elements by mass transfer
from the primary AGB star. This means that the original AGB
envelope must have been even more enriched in low-mass s-elements, by
a factor of at least 0.2 dex. The present sample of C-stars
as well as that of similar $s$-enriched stars is not an ample one and the
average values should be taken with some caution. Furthermore, both
the MS/S and the C-stars show more complex spectra crowded by
molecular lines than the warmer Ba~II giants, which
are also less C-rich. Summing up, we shall have to wait for more ample
statistics and refined model atmospheres of the C-rich stars to
make further inferences in these comparisons.

Three stars in the sample, V CrB, VX Gem  and IY Hya, deserve
special attention. The first two stars are Mira variables with
periods of 357 and 379 days, respectively. 
For the Mira variable V CrB, \citet{kip98}
derived [Fe/H] = $-$2.12 and [ls/Fe] $\approx$ +1.5 and
concluded that the abundance pattern in this star is similar to
that of other late-type CH-stars. The values we derived in this star
for these figures are very different (see Table~2). We do not
have an easy explanation for this discrepancy, as we used the
same model atmosphere (kindly provided by T. Kipper).
Because of $T_{\rm{eff}}$ variations in Mira variables, we have
analyzed this star using other $T_{\rm{eff}}$ values as proposed
in the literature (see discussion in Kipper's paper). However,
the best fit to the observed spectrum is obtained with the value
adopted here, $T_{\rm{eff}} = 2250$ K. Inspection of Figure 3 in
\citet{kip98} makes clear that the heavy element
abundances derived by spectral synthesis just fit the cores of the
absorption lines without reproducing consistently the global
behavior of the observed spectrum. We thus believe that the line
list used in Kipper's analysis of V CrB does not contain
important CN and C$_2$ absorption, what unavoidably leads
to the derivation of larger [X/Fe] ratios. This could be the reason for the
important differences between the two analyses.

Also for the Mira variable VX Gem the spectroscopic
abundance analysis is rather complex. This star
shows line-doubling in the stronger lines (Na I D doublet,
7698 {\AA} K I, 6708 {\AA} Li I). The hydrogen lines
($\alpha, \beta, \gamma$) as well as the 5980 {\AA} Mg I line
are in emission. The Ca II infrared triplet shows P-Cygni
profiles. All these features indicate the existence of a
circumstellar envelope. On the other hand, $T_{\rm{eff}}$
for this star is rather uncertain. We have adopted 2500 K
according to the analysis of \citet{gro89}, but \citet{ohn96}
derived 3050 K from infrared photometry and classified this star
as a SC-star. Indeed, effective temperatures in Mira variables
change during the pulsational phase. However, we have found a
better agreement between the theoretical and the observed spectrum
in all the spectral ranges analyzed when adopting $T_{\rm{eff}} = 2500$
K. The abundances derived in this star therefore have to be
considered with caution. In fact, although the Rb I line is
clearly visible in its spectrum we prefer not to estimate the
[Rb/M] ratio because of the probable contamination with a
circumstellar Rb line.

The situation concerning IY Hya is even more confusing. This star
is a very red object with a high mass-loss rate, $4\times
10^{-6}$ $M_\odot$ yr$^{-1}$ \citep{jur89}. Furthermore, IY Hya
is one of the few Galactic super Li-rich C(N)-stars, with log (Li/H)
$+~12 \sim 4.0$ \citep{abi99}. The atmosphere parameters for this
star are quite uncertain since no direct determination exists in
the literature. Evidence of this is the fact that we do not find
a unique CNO abundance choice (within error bars) that would give a good
fit for all the spectral ranges analyzed here. Thus, because of
the very uncertain abundance ratios derived in V CrB, VX Gem  and
IY Hya, we have plotted their abundance ratios with smaller
symbols in the figures below, to indicate that they have a lower
weight in our analysis.

\subsection{Technetium}

As an additional hint to the abundance pattern found in our stars
we searched for the presence of technetium. The search was
based mainly on the analysis of the intercombination Tc I line at
5924.47 {\AA} (see Paper I for a detailed explanation
of the procedure followed in the analysis of this Tc blend).
Unfortunately, the stronger resonance Tc lines around 4260 {\AA}
are inaccessible in most C-stars because of the strong flux
depression in these stars below $\sim$ 4400 {\AA}. In the
last column of Table 2 we have indicated whether or not Tc is
identified for each star. A ``yes'' entry means that the best fit
to the 5924 {\AA} blend is obtained when a Tc abundance different
from zero is used in the spectral synthesis\footnote{Examples of
the quality of the fits to the Rb and Tc blends can be appreciated
in the corresponding figures shown in Papers I and II.}. With a
``no'' entry we mean that a synthetic spectrum with no-Tc does
not significantly differ from another one computed with a very
small Tc abundance. In some stars the best fit to the Tc blend is
obtained using a small Tc abundance but not strictly zero. These
stars have been quoted with a ``doubtful'' (dbfl) entry since in
our judgment the evidence favours the absence of Tc. We have to
note however, that a ``no'' in Table 2 does not necessarily mean
that Tc is absent since, as the 5924 {\AA} Tc I line is much
weaker and in a more crowded zone than the resonance lines near
4260 {\AA}, a detailed study of the latter might eventually
reveal the presence of Tc. Nevertheless, in two stars for which
it was possible to analyze both the 5924 {\AA} and 4260
{\AA} spectral ranges (VX Gem and W Ori), we reached the same
conclusion for both lines (see Table 2). If we exclude the stars quoted as
doubtful, 50\% of the stars in Table 2 show Tc, the others are
compatible with a non-detection. Interestingly enough, the same
figure is found in MS/S stars with/without Tc \citep{smi85,smi86,smi90}.
If one applies the
same rules currently accepted for the MS/S giants to our C-rich objects, then
Tc-no C(N)-stars would be extrinsic AGBs (candidates to have a WD
companion), with extra dilution
of $s$-element abundances related to mass transfer from a companion.
However, we do not find any correlation
between the level of heavy-element enhancements and the presence
of Tc. In fact, for the Tc-yes and Tc-no stars we found
$<$[ls/Fe]$>$ = +0.53 and $<$[ls/M]$>$ = +0.48, respectively. 

As far as we know, there is no evidence of binarity for any of the stars
studied here; despite this, we can take a further step in order to
identify the extrinsic (binary) or intrinsic nature of our
C-stars by studying their IR colours in the same way as 
\citet{jor93} did for S-stars. Infrared colours provide a measure
of the optical properties of circumstellar dust. Such dust is
expected to be around intrinsic (non-binary) C-stars, and thus the
presence or absence of dust emission might enable us to discriminate between
intrinsic and extrinsic C-stars. Following the results of \citet{jor93} most Tc-rich S
stars have IR excesses ($R\ga 0.1$), which would indicate that
they owe their chemical peculiarities to intrinsic nucleosynthesis
processes. On the contrary, Tc-deficient (binary) S stars usually
show $R<0.1$.  To estimate this IR excess due to
dust, the flux ratios $R=F(12\mu\rm{m})/F(2.2\mu\rm{m})$ were
computed. The 2.2 $\mu$m infrared flux was derived from the K
magnitude taken from the Two Micron Sky Survey \citep{neu69, cla87} and the 12 $\mu$m fluxes from the IRAS point source
catalogue. The small sample of C-stars and the fact
that our Tc analysis is based on a weak line in a crowded zone
prevents us from reaching any firm conclusion. However, we find
that 72$\%$ of Tc-no stars in Table 2 have $R<0.1$, while this
figure decreases to $45\%$ in the Tc-yes stars. This may be an
indication of some correlation in the same direction as found in S
stars. A similar study in a larger sample of C-stars is clearly
necessary.

\section{The stellar and nucleosynthesis reference models}

In order to compare our observed abundances of C(N)-stars with
predictions from current models of AGB evolution and
nucleosynthesis, we use the $s$-process nucleosynthesis results
from Gallino et al. (1998), Travaglio et al. (2001) and Busso et al.
(2001), which are based on a large set of
AGB stellar models, for stars of initial mass between 1.5 and 7 \ms,
made with the FRANEC evolutionary code
(Straniero et al. 1995, 1997, 2001). The $s$-process
nucleosynthesis calculations were performed with a post-process
code and were extended over a wide range of metallicities.
Specifically, we make use here of the models for 1.5 and 5 \msb
stars, in order to show how different predictions about
abundances and abundance ratios at the Sr-Y-Zr peak that can be
obtained when the initial mass is assumed to belong to the LMS or
to the IMS range. Reimer's (1975) parameterization for mass
loss was adopted ($\eta$ = 0.7 and 10, for the 1.5~\msb and the 
5~\msb cases, respectively). When models are run at nearly solar
metallicity, the one relevant for most of our sample stars,
third dredge-up is found to start after a few thermal pulses.
The parameter $\lambda$ (defined as
$\Delta$$M$(TDU)/$\Delta$$M_{\rm H}$, where $\Delta$$M$(TDU) is
the mass dredged-up in one mixing episode and $\Delta$$M_{\rm H}$
is the extension in mass by which the H shell advances in an
interpulse period) is on average 0.24 and 0.38 in the two cases.
TDU is followed in detail pulse after pulse, mixing with the
envelope material from the He intershell rich in \cdb and $s$-elements.
In this way we can follow the stellar
envelope composition evolving in time while the star ascends the
TP-AGB phase.

The activation of the $^{13}$C neutron source requires some protons from the
envelope to penetrate into the He intershell when TDU is
established. Subsequently, proton captures on the abundant
$^{12}$C allow the formation of a
small $^{13}$C-$pocket$ in the uppermost layers of the He
intershell. All $^{13}$C nuclei are then consumed by
$\alpha$-captures during the interpulse phase, while the intershell
zone is in radiative conditions \citep{str97}. Successful models
of the \ct-pocket formation have been presented in the recent
years, as a consequence of the application of diffusive or
hydrodynamical simulations of the partial mixing occurring at the
H-C interface (Iben \& Renzini 1982; Hollowell \& Iben 1988;
Herwig et al. 1997; Cristallo et al. 2001). A general consensus is
emerging that such partial mixing events do indeed
occur: they might be stimulated by rotational shear
(Langer et al. 1999). However, the above models are still very
speculative and cannot be used to make general predictions
suitable to perform a comparison with observations. Therefore,
the amount of $^{13}$C in the pocket and the distribution
within it must still be considered as free parameters.
Following the approach by Busso et al. (2001), we chose here to
adopt a wide range of \ctb production in
the pocket. We start from the reference choices made in Gallino
et al. (1998) for LMS, and in Travaglio et al. (2001) for IMS:
both choices were called {\it standard} (ST) in the original papers. They
correspond to a total \ctb mass of 4 $\times$ 10$^{-6}$ \msb in
LMS and of 4 $\times$ 10$^{-7}$ \msb in IMS. 
The decrease by an order of magnitude of the strength of the  $^{13}$C neutron 
source in the pocket corresponds to the overall decrease by the same order of the 
mass of the He intershell, the TDU efficiency, and the interpulse period.
Then we scale such abundances upward and
downward, obtaining cases with efficiencies from
a factor of two higher (ST$\times$2) to a factor of 12 lower
(ST/12).

The envelope composition is progressively enriched by TDU in C and
$s$-elements, reaching the condition for a C(N) star (C/O $>$ 1) 
in the last TPs. For AGB stars of initial mass 1.5 \msb this occurs for any 
metallicity up to [Fe/H] = +0.2. Table 5 shows the predicted C/O 
and \cd/\ctb ratios in the envelope after the last TDU episodes where the 
condition C/O~$>$~1 is met for AGB models of 1.5, 3 and 5 \msb and
for the three metallicities [Fe/H] = $+0.2$, 0.0 and $-0.3$. 
For each mass, in the third column we also report the remaining envelope
mass at the given TP. With the adopted mass loss prescriptions, at 
[Fe/H]~=~+0.2 we do not find C/O $>$ 1 in the envelope 
of AGB stars of 3 and 5 \ms, although we are not far from
reaching this condition. Our sample stars span a range $-0.3$ to +0.3 in
[Fe/H], but the estimated uncertainty of the metallicity and, even more,
the observational uncertainty in the mass loss rate does not allow
us to make a more detailed analysis such as, for example, to exclude on
that all stars with nominal [Fe/H] higher than solar are 
of initial mass below 3~\ms. 

\section{Discussion}

\subsection{The neutron density and the mass of C stars}

Figures 1 and 2 show the run of observed abundance ratios
[X/M], for Rb, Sr, Y and Zr, compared with AGB model predictions
for stars of initial
mass 1.5 and 5 \msb and different metallicities. The plotted
curves pick up the envelope composition when the photospheric C/O
ratio reaches unity, because we must fulfill the general
requirement that our stars are carbon rich with C/O ratios
in the range 1.01 to 1.20. The $s$-element abundance predictions for AGB
stars of $M= 3$ $M_\odot$ are almost identical to the $M = 1.5$ $M_\odot$ case.
An exception is [Rb/Fe], which is 0.1 dex higher for all choices of the \ct-pocket.
Different mass loss prescriptions or details on TDU
pulse after pulse would not affect much the results, provided C/O = 1 is reached in 
the envelope. 

Table 6 gives the Rb, Sr, Y, and Zr expectations at C/O =1 for the various AGB
models and different \ct-pocket prescriptions. All sample stars are
reproduced by our theoretical models in the range ST/1.5 to ST/3. 
Figures 1 and 2 show
only a weak tendency of the curves to be non-monotonic versus [M/H]
in the range of the bona fide observed C(N) stars.
For the 5 \msb model a decrease in $s$-element abundances with
decreasing metallicity appears.
This predicted trend of envelope abundances is
quite different from what has recently been shown in \citet{bus01}
for the same AGB stellar models, where in particular a steep increase
of [ls/Fe] resulted with decreasing [Fe/H], up to a maximum at around +1.2,
(which is reached at [Fe/H] $\sim$ $-0.5$ for the ST case), followed by a decrease 
at lower metallicities. The discrepancy is only apparent and deals with the 
particular constraint to be meet here, i.e. with the mentioned fact
that in our sample C(N) stars, C/O is close to 1 (see Table 1). The trends shown in
Busso et al. (2001) are instead representative of the last
occurrence of TDU in the models. For a 1.5 \msb star, in our
calculations this corresponds to the 17th TDU episode; for solar
metallicity this implies C/O = 1.44. 

In AGB stars of lower metallicity, given that the \cdb
production in the top layer of the He intershell is always close to 0.2 in mass fraction,
the carbon enrichment in the envelope is more easily established. 
Therefore the envelope C/O ratio increases steadily with decreasing 
metallicity, and is moderated only by the fact that
at low [Fe/H] the stars condense from an 
O-enhanced interstellar medium \citep{whe89}. At [Fe/H] = $-$1.3
(close to the estimated metallicity of V CrB) in a 1.5 \msb model 
after 17 TDUs, C/O = 11.4 and \cd/\ctb = 2221.
For such a metallicity, a C/O ratio close to unity is
found at the very early occurrence of TDU (with a corresponding \cd/\ctb =
135), when the $s$-process
abundances have been barely affected by neutron captures in the
He intershell. We underline that this finding should be
considered as a general rule for C(N) stars, as observations always reveal a
moderate C enrichment. This may imply some selection effect:
intrinsic AGB C-stars with higher C/O ratios should exist, as
high values of C/O have been observed in both post-AGB stars
reaching the pre-planetary nebula stage (Reddy, Bakker, \&
Hrivnak 1999) and in extrinsic C-rich objects originated by mass
transfer phenomena (Vanture 1992c). However, an excess of carbon is
immediately translated into a copious production of carbon-rich
dust and into a high opacity of the circumstellar environment, to
the point of hiding the photosphere at optical wavelengths. This
is the case of some infrared carbon stars observed 
by IRAS (van Loon et al. 1998; Marengo et al. 1999). Therefore, unlike 
MS-S giants or the extrinsic Ba stars, which display
different levels of C enrichment, the {\it visible} C(N) stars represent a sort of
snapshot taken at a particular moment during the  TP-AGB evolution,
always representing the first appearance of a carbon-rich
composition. 

Figures 1 and 2 show, however, that non-negligible differences in the $s$-process
predictions at C/O =1 from a LMS and from an IMS exist. The IMS models show a
considerably narrower strip of predicted abundances when the amount of \ctb burnt 
in the pocket is varied as compared with the LMS case. This is true especially 
for [Rb/Fe]. These IMS predictions, as a whole, seem scarcely compatible with
most sample stars. 

In Figures 3 and 4 we plot the elemental ratios [Rb/Sr], [Rb/Y] and [Rb/Zr]
versus metallicity. Of the three panels in the figures, the one
yielding the clearest distinction is the second one, [Rb/Y].
According to the discussion of \S~2, the fact that
IMS models (where the $^{22}$Ne source is active) compare less
favourably with the data is an indication that most
neutron fluxes at the origin of the observed $s$-element
enhancements come from the \ctb source, and consequently that 
most C(N) stars are of low initial mass.

Eventually, in Figure 5 we use [ls/M] as the average abundance of [Y/M] and [Zr/M], 
after excluding Sr, whose abundance is more uncertain. 
The entire sample of stars having a measurement for 
the three elements: Rb, Y and Zr, are (with the exception of IY Hya)
compatible with a metallicity [M/H] = 0.0 $\pm$ 0.30 and actually
fulfills this requirement within 1 $\sigma$.
Figure 5 considers only model sequences above [Fe/H] = $-$0.3.
Along the theoretical curves, the predicted [Rb/Y] increases with increasing
[Fe/H]; exception is the ST
case for $M$ = 1.5 \ms, for which the lower [Fe/H] corresponds to
the lower [ls/Fe].
The plot immediately displays the characteristics
discussed above: LMS models occupy the region of negative [Rb/Y],
IMS models stay on the opposite side, and the partial overlapping
of figures 1 and 2 is avoided. 

\subsection{Carbon isotopes and related problems}

Figure 6 shows in graphic form the C/O observed ratio versus \cd/\ct.
Current stellar models yield a small reduction in C/O (by 30~$\%$) 
and a consistent reduction of the \cd/\ctb ratio (by a factor 4) during
the first ascent of the red giant branch, caused by the first
dredge up mechanism (FDU). The reduction by a factor of 4 from the solar ratio (\cd/\ctb = 89) 
is obtained after decreasing by one third of the initial $^{12}$C abundance and through an increase 
by a factor $\sim$2.5 of $^{13}$C, both changes being due to proton captures on $^{12}$C
during the main sequence in the deep regions of the radiative H-rich
envelope (e.g., Dearborn 1992). Note that the resulting ratio after the FDU is independent of
the initial \cd/\ctb value. The same FDU results are obtained for different
initial metallicities, even when one accounts for the fact that, in Galactic evolution,
$N(^{12}$C) scales as $N(Fe)$, while $N(^{13}$C) scales as $N(Fe)^2$ (Iben 1977; Boothroyd \& Sackmann 1988).
 
However, at least for masses smaller than about 2 \msb the canonical reduction of the 
$^{12}$C/$^{13}$C ratio at FDU down to $20-25$ is insufficient to
explain the observations in red giants (Gilroy \& Brown 1991;
Charbonnel, Brown, \& Wallerstein 1998; Gratton et al. 2000).
A similar discrepancy with model predictions is shown by oxygen and carbon isotopes
measured in dust grains of clear AGB circumstellar origin,
recovered from meteorites. This is true both for C-rich (Anders
\& Zinner 1993) and for O-rich grains (Nittler et al. 1994; Hutcheon et al.
1994). Explanation of this evidence involves 
a relatively slow process of mixing occurring during the RGB in the
radiative region located between the bottom of the
convective envelope and the H-burning shell, perhaps induced by rotation
(Sweigart \& Mengel 1979). This process is sometimes called
Cool Bottom Process, or CBP (Wasserburg, Boothroyd, \&
Sackmann 1995). CBP has been shown to yield quite naturally
enhancements of $^{13}$C and of $^{17}$O and reductions in
$^{18}$O. Several phenomenological models have been presented for such
phenomena (Boothroyd, Sackmann, \& Wasserburg 1994;
Denissenkov \& Weiss 1996; Weiss, Denissenkov, \& Charbonnel 2000). The general
agreement with observations has led to acceptance
that such circulations should indeed be active, thus further reducing in LMS  
red giants the \cd/\ctb ratio established after the FDU.
The exact extent of the depletion as a function of mass for
each metallicity is, to date, rather uncertain, as the
requirement of sampling low mass objects has often led to obtain the
measurements in metal poor clusters, so that the two parameters
(mass and metal content) are coupled and difficult to separate.
Recently, Charbonnel, Brown, \& Wallerstein (1998) have
started a project on field stars whose parallax was accurately
measured by HIPPARCOS, and a clearer view is now emerging, in
which giant stars of the old Galactic disc population (at
metallicities [Fe/H] $\sim$ $-$0.6) can reach \cd/\ctb ratios
close to 7. For more metal-rich stars a reasonable estimate might
be $10-15$ (Gilroy \& Brown 1991).
Not having a suitable model for CBP, we chose this
estimate as the \cd/\ctb ratio present in early AGB phases of
LMS before the onset of thermal pulses: the adopted tentative value 
is 12. For stars more massive that $\sim 2.5$ \ms, we assumed \cd/\ctb~=~24.

When we then add \cdb to the envelope by TDUs along the AGB,
the isotopic ratio raises. At C/O$\sim 1$ a pre-AGB value of around 25 would
translate into a final value a bit higher than the solar isotopic ratio ($^{12}$C/$^{13}$C=89). 
Indeed, after a few TPs, when TDU 
is operating, a nearly constant mass fraction of \cdb, $X(^{12}\rm{C})\sim 0.23$, is built in the He intershell
by partial He burning in the TP. This is independent of the core mass, i.e. on
the AGB initial mass. About identical results are obtained by different authors
(see e.g., Iben 1976, 1982; Boothroyd \& Sackmann 1988; Straniero et al. 1995, 1997; Mowlavi 1999).
Only a marginal production of $^{16}$O occurs in the TP, so that the oxygen
abundance in the envelope is not affected by TDU. Also the \ctb abundance
in the envelope is not affected by TDU, since no \ctb nuclei are left in the He
intershell, while the mass of the He intershell cumulatively mixed with
the surface is a small fraction of the envelope. Under those conditions, 
it is easy to evaluate the \cd/\ctb in the envelope at C/O = 1. It
results: \cd/\ctb  $\sim$ 3.6~$\times$~(\cd/\ctb)$_{\rm FDU}$. This relation is independent of
metallicity if both \cdb and $^{16}$O are assumed to scale as Fe.
Actually, for [Fe/H]~=~$-0.3$ we introduced a modest $\alpha$-enhancement in
the initial $^{16}$O abundance with respect to a solar-scaled metallicity
composition, according to an average observational trend [O/Fe] =
$-0.5$~[Fe/H] (cf. Wheeler, Sneden, \& Truran 1989). This implies
an increase by about 30$\%$ of the predicted $^{12}$C/$^{13}$C ratio at C/O~=~1.

As shown in figure 6, only two stars, V Aql and  W Ori, show the expected ratio
at C/O $\sim$ 1, without requiring any CBP to operate.
Those two stars should have a mass higher than 2.5 $M_\odot$, perhaps around 3 \ms.
When we consider the majority of our sample stars, which have
previously been shown to be of low mass, we immediately see that
CBP must have reduced the \cd/\ctb ratio. Actually, even a CBP operating
only on the RGB would be insufficient, as according to the previously
given rule, starting from ($^{12}$C/$^{13}$C$)_{\rm{FDU}}\sim 12$ we would have final values of 
43 (see Figure 6). Hence CBP should be active also later, in the early
AGB phase, to operate changes larger than we assumed and thus explain the later
part of the $^{12}$C/$^{13}$C distribution. The possibility exists that CBP operates
also in the TP-AGB phase, though calculations have not been performed yet. 
A different efficiency of rotationally induced meridional mixing 
(Sweigart \& Mengel 1979) in different stars might also accomplish this job.

Furthermore, in the above discussion we assumed that all sample stars are intrinsic 
AGBs. This actually may not be true. A considerable fraction of C(N) stars might be extrinsic AGBs 
belonging to binary systems (see $\S~4.1$). In this case, the observed star could be ascending  
the early AGB phase, having being enriched in carbon and $s$-process 
elements by mass transfer from the more massive companion while this last was on the
AGB. In this case FDU on the companion star can further reduce the \cd/\ctb ratio
\footnote{Note, that actually there is a controversy on the typical $^{12}$C/$^{13}$C
ratio in C(N)-stars (see Lambert et al. 1986; Ohnaka \& Tsuji 1996-98; de Laverny \&
Gustafsson 1998-99; Sch\"oier \& Olofsson 2000).}.

\section{Conclusions}
 
In this paper we have presented new high-resolution spectroscopic
measurements for carbon-rich AGB stars belonging to the C(N) type.
We have focused our attention on $s$-elements around the
$^{85}$Kr reaction branching of the $s$-process path, showing
that the comparative analysis of Rb, Sr, Y, and Zr yields information on the 
neutron density prevailing during the neutron capture nucleosynthesis processes
occurring in the He intershell. Despite the large uncertainty existing 
in the derivation of element abundances in these stars, comparison with AGB
nucleosynthesis models, which were already capable of explaining
the abundances in other intrinsic and extrinsic classes of AGB
stars, confirms a general agreement between theoretically
predicted photospheric compositions and observations. The main
implication of such an analysis is that most C(N) stars are of low mass, 
experiencing $s$-process
nucleosynthesis phenomena dominated by the neutron source
provided by $\alpha$ captures on \ctb in radiative intershell
layers. When the
\cd/\ctb ratios are considered, their generally low values
require the activation of slow circulation mechanisms below the
formal convective border of the envelope, bringing material close
to the H-burning shell and therefore increasing the abundance of
\ct. These mechanisms are indeed known to occur in stars below
about 2 \ms: this implies, therefore, the conclusion that the
majority of C(N) stars are of small mass. Exceptions might
exist in a very few cases, though this speculation is highly uncertain and 
should be verified by
more extensive measurements of $s$-process elements, especially
at the Ba-peak.

\acknowledgments We thank the referee for the very careful reading of the
manuscript and his valuable comments and suggestions. C. Abia is grateful to P. de Laverny and B. Plez
for providing the CN and C$_2$ line lists in the wavelength
ranges studied. Data from the VALD database at Vienna were used
for the preparation of this paper. K. Eriksson, and the stellar
atmosphere group of the Uppsala Observatory are thanked for
providing the grid of atmospheres. The 4.2m WHT is operated on
the island of La Palma by the RGO in the Spanish Observatory del
Roque de los Muchachos of the Instituto de Astrof\'\i sica de
Canarias. This work was also based in part on observations
collected at the German-Spanish Astronomical Centre, Calar Alto,
Spain. It was partially supported by grants PB96-1428, ESP98-1348 and
AYA2000-1574, by the Italy-Spain agreement HI1998-0095 and by the
Italian MURST-Cofin2000 project `Stellar Observables of
Cosmological Relevance'.

\clearpage

\begin{deluxetable}{lcccccc}
\tablewidth{410pt}
\tablecaption{Data for Program Stars}
\tablehead{
\colhead{Star}           & \colhead{Spec. Type} &
\colhead{$T_{\rm eff}$ }& \colhead{Ref.\tablenotemark{a}} &
\colhead{C/O}           & \colhead{$^{12}$C/$^{13}$C}     &
\colhead{Ref.\tablenotemark{b}}}
\startdata
AQ And & C5,4 & 2970$^\ast$& 2 &1.02 & 30 & 1\\
AW Cyg & C3,6 & 2760       & 2  & 1.03 & 21 & 2 \\
EL Aur & C5,4 & 3000       & 6  &1.07 & 50 & 2 \\
HK Lyr & C6,4 & 2866       & 2  &1.02 & 10 & 2 \\
IRC $-$10397 & N & 2600      & 6  &1.01 & 20 & 2 \\
IY Hya &  N   & 2500       & 6  & 1.02 & 15 & 2 \\
LQ Cyg & C4,5 & 2620       & 2  &1.10& 40 & 2\\
RX Sct & C5,2 & 3250       & 2  &1.04 & 60 & 2 \\
S Sct & C6,4  & 2895       & 1  &1.07 & 45 &3 \\
SY Per & C6,4 & 3070       & 3  &1.02 & 43 & 2 \\
SZ Sgr & C7,3 & 2480       & 2  &1.03 & 8 & 2\\
TY Oph & C5,5 & 2780$^\ast$& 2  &1.05 & 45 & 2 \\
U Cam  & C3,9 & 2670       & 2  &1.02 & 40 & 2 \\
UU Aur & C5,3 & 2825       & 1  &1.06 & 50 & 3\\
UV Aql & C6,2 & 2750       & 2  &1.005 & 19 & 2 \\
UX Dra & C7,3 & 2900       & 1  &1.05 & 26 & 2\\
V Aql & C6,4 &  2610        & 1  &1.16 & 90 & 2 \\
V CrB & C6,2 &  2250        & 4  &1.10 & 10 & 4\\
V Oph & C5,2 &  2880$^\ast$ & 2  &1.05 & 11 & 1 \\
VX Gem& C7,2 &  2500        & 5  &1.01 & 60 & 2 \\
V460 Cyg & C6,3& 2845     & 1  &1.06 & 61 & 4\\
V781 Sgr &  N &  3160      & 2  &1.02 & 35 & 3\\
W Ori & C5,4 & 2680        & 1  &1.20 & 79 & 4\\
WZ Cas& C9,2J& 3140        & 3  & 1.005 & 4  &3\\
Z Psc & C7,2 & 2870        & 1  & 1.01 & 55 & 4 \\
\tablenotetext{a}{References for $T_{\rm {eff}}$: (1) Lambert et al. (1986); (2) Ohnaka \& Tsuji (1996);
(3) Dyck, van Belle \& Benson (1996); (4) Kipper (1998); (5) Groenewegen et al. (1998); (6) For these stars
we have estimated the effective temperature by comparing their spectra  with the temperature sequence
spectral atlas for C-stars created by Barnbaum et al. (1996).}
\tablenotetext{b}{Source for the $^{12}$C/$^{13}$C ratio: (1) Ohnaka \& Tsuji (1996);
(2) Derived in this work; (3) Lambert et al. (1986); (4) Kipper (1998)}
\enddata
\end{deluxetable}

\clearpage

\begin{deluxetable}{lcccccc}
\tableheadfrac{0.08}
\tablecolumns{7}
\tablewidth{0pt}
\tablecaption{Abundances Derived in  Program Stars}
\tablehead{
\colhead{Star}                    & \colhead{[M/H]}    &
\colhead{[Rb/M]}                  & \colhead{[Sr/M]}   &
\colhead{[Y/M]}                   & \colhead{[Zr/M]}& \colhead{Tc}}
\startdata
AQ And   &  0.02  & 0.3  & 0.6 & 0.7 & 0.5& no \\
AW Cyg   &  0.0   & 0.2  & 0.4 & 0.3 & 0.5& yes \\
EL Aur   &  $-0.06$ & 0.4 & 0.5 & 0.7 & 0.8& no \\
HK Lyr  & $-0.10$  & 0.2 & 0.3 & 0.7 & 0.7&yes \\
IRC $-10397$ &  0.10 & 0.1  & 0.3 & \nodata & \nodata&\nodata \\
IY Hya  & $-0.80$  & 0.5 & 0.6 & 0.6 & 1.0&\nodata \\
LQ Cyg   &   0.25 & 0.2  & 0.5 & 0.8 & 0.7&no \\
RX Sct  & $-0.05$ & 0.4 &  0.4 &  0.7 & 0.8& no \\
S Sct   & 0.01 & 0.5 & 0.8 & 0.7 & 0.5&yes \\
SY Per & $-0.30$ & 0.1  & 0.5 &  \nodata & \nodata&doubf \\
SZ Sgr & $-0.04$ & 0.1 &  0.4 & 0.8 & 0.8&doubf\tablenotemark{a} \\
TY Oph &  0.10 & 0.2 & 0.6 & 0.7 & 0.6&doubf \\
U Cam &  $-0.09$ & \nodata & \nodata & 0.6 & 0.4&no \\
UU Aur &  0.06 & 0.4 &   0.4 & 0.6 & 0.6&no \\
UV Aql & $-0.06$ & 0.6 & 0.5 & 0.7 & 0.9&yes \\
UX Dra & $-0.20$ & 0.0  & 0.8 & 0.8 & 0.8&yes\tablenotemark{a} \\
V Aql & $-0.05$ & 0.3 & 0.5 & \nodata &\nodata&no \\
V CrB & $-1.35$ & $-0.2$  & 0.3 & 0.4 & \nodata&no\tablenotemark{a} \\
V Oph & 0.0 &  $-0.2$ & 0.2 & 0.3 & 0.4&\nodata \\
VX Gem & $-0.15$ &\nodata & 0.6 & 0.5 & 0.6&yes\tablenotemark{a} \\
V460 Cyg & $-0.04$ & 0.4 & 0.2 & 0.4 & 0.4&yes \\
V781 Sgr & 0.10 & 0.1 & 0.5 & 0.3 & 0.2&yes \\
W Ori &  0.05 & \nodata & 0.0 & 0.3 & 0.1&no\tablenotemark{a} \\
Z Psc & $-0.01$ & 0.9 & 0.9& 1.0 & 0.8&yes \\
\tablenotetext{a}{Little et al. (1987) classified SZ Sgr as {\it possible} star showing Tc and
reported Tc detection in UX Dra. Kipper (1998) sets an upper limit to the Tc abundance in
V CrB (log (Tc/H) $+12$ $\la$ $-0.5$) from the analysis of the 5924 {\AA} Tc I line. However, our analysis is
compatible with no Tc. For VX Gem and W Ori we were able to get spectra in the 4260 {\AA} region
whose analysis confirmed the figure found from the 5924 {\AA} Tc I analysis.}
\enddata
\end{deluxetable}
\clearpage

\begin{deluxetable}{cccccccccccc}
\tablenum{5}
\scriptsize
\tablecolumns{12}
\tablewidth{0pt}
\tablecaption{C/O and $^{12}$C/$^{13}$C Predictions in the Envelope}
\tablehead{
\colhead{} & \multicolumn{3}{c}{1.5 $M_\odot$} & \colhead{} & \multicolumn{3}{c}{3 $M_\odot$} & \colhead{} & \multicolumn{3}{c}{5 $M_\odot$}\\
\colhead{Metallicity} &\colhead{C/O} & \colhead{$^{12}$C/$^{13}$C} & \colhead{$M_{\rm{env}}$} & \colhead{} & \colhead{C/O} & 
\colhead{$^{12}$C/$^{13}$C} & \colhead{$M_{\rm{env}}$} & \colhead{} & \colhead{C/O} & \colhead{$^{12}$C/$^{13}$C} & \colhead{$M_{\rm{env}}$} }
\startdata
[Fe/H]$=+0.2$&1.00 & 43 & 0.28 & &\nodata & \nodata & \nodata & &\nodata & \nodata & \nodata \\
&1.05 & 45 & 0.24 & &\nodata & \nodata & \nodata & &\nodata & \nodata & \nodata \\
\cline{1-12}
[Fe/H]$=+0.0$ & 1.06  & 46 & 0.38 & & 1.00 & 91 & 1.05 & & 1.00 & 92 & 1.23 \\
             & 1.17  & 51 & 0.34 & & 1.04 & 94 & 0.98 & & 1.04 & 95 & 1.16 \\
             & 1.27  & 55 & 0.31 & & 1.08 & 98 & 0.91 & &1.08 & 99 & 1.09 \\
             & 1.37  & 60 & 0.28 & & 1.11 & 101& 0.83 & &1.12 &103 & 1.02\\
             & 1.45  & 63 & 0.24 & & 1.14 & 104& 0.76 & &1.16 &107 & 0.95\\
             &\nodata&\nodata&\nodata&&\nodata&\nodata&\nodata& &1.21 & 111 & 0.88\\
             &\nodata&\nodata&\nodata&&\nodata&\nodata&\nodata& &1.26 & 116 & 0.80\\
             &\nodata&\nodata&\nodata&&\nodata&\nodata&\nodata& &1.32 & 122 & 0.73\\
\cline{1-12}
[Fe/H]$=-0.3$&1.01 & 62 & 0.44 & & 1.01 & 131 & 1.31 & & 1.01 & 132 & 1.58 \\
             &1.16 & 72 & 0.41 & & 1.08 & 140 & 1.24 & & 1.05 & 137 & 1.51 \\
             &1.31 & 81 & 0.38 & & 1.13 & 146 & 1.18 & & 1.09 & 143 & 1.44 \\
             &1.47 & 92 & 0.34 & & 1.18 & 154 & 1.11 & &1.13 & 149 & 1.37\\
             &1.61 &101 & 0.31 & & 1.23 & 160 & 1.05 & &1.18 & 155 & 1.30\\
             &1.76 &110 & 0.28 & & 1.28 & 167 & 0.98 & &1.23 & 162 & 1.23\\
             &1.86 &117 & 0.24 & & 1.34 & 174 & 0.91 & &1.28 & 169 & 1.16\\
             &\nodata&\nodata&\nodata& &1.38 & 180  & 0.83 && 1.33 & 176 & 1.09\\
             &\nodata&\nodata&\nodata& &1.44 & 187  & 0.76 && 1.39 & 184 & 1.02\\
             &\nodata&\nodata&\nodata&&\nodata&\nodata&\nodata&&1.45&193&0.95\\
             &\nodata&\nodata&\nodata&&\nodata&\nodata&\nodata&&1.52&202&0.88\\
             &\nodata&\nodata&\nodata&&\nodata&\nodata&\nodata&&1.59&212&0.80\\
\enddata
\end{deluxetable}
\clearpage

\begin{deluxetable}{lccccccc}
\tablenum{6}
\scriptsize
\tablecolumns{8}
\tablewidth{480pt}
\tablecaption{Abundance Ratios at C/O$=1$ for different AGB models
and $^{13}$C-pocket prescriptions}
\tablehead{
\colhead {Model}&\colhead{$^{13}$C-pocket}&\colhead{C/O}&\colhead{$^{12}$C/$^{13}$C}&\colhead{[Rb/Fe]}&\colhead{[Sr/Fe]}&\colhead{[Y/Fe]}&\colhead{[Zr/Fe]}}
\startdata
1.5 $M_\odot$ [Fe/H]$=+0.2$& ST&1.00 & 43 & 0.51 & 0.91 & 0.82 &0.70 \\
&ST/1.5& 1.00 & 43   & 0.31 & 0.61 & 0.52 & 0.39 \\
&ST/3& 1.00 & 43 & 0.09 & 0.16 &  0.10& 0.05 \\
\cline{1-8}
1.5 $M_\odot$ [Fe/H]$=+0.0$& ST  & 1.06 & 46 & 0.56 & 1.06 & 0.96 & 0.82 \\
 &ST/1.5 & 1.06 & 46 & 0.37 & 0.75 & 0.65 & 0.54 \\
 &ST/3 & 1.06 & 46 & 0.12 & 0.28 & 0.20 & 0.12 \\
\cline{1-8}
1.5 $M_\odot$ [Fe/H]$=-0.3$&ST& 1.07 & 46 & 0.39 & 1.12 & 1.12 & 1.08 \\
&ST/1.5 & 1.07 & 46 & 0.37 & 0.95 & 0.86 & 0.73 \\
&ST/3& 1.07 & 46 & 0.16 & 0.46 & 0.37 & 0.29\\
\cline{1-8}
3 $M_\odot$ [Fe/H]$=+0.0$&ST    &1.00& 91  & 0.71  & 1.05& 0.98 & 0.85\\
                         &ST/1.5&1.00& 91  & 0.51  & 0.75& 0.68 & 0.57 \\
                         &ST/3  &1.00& 91  & 0.20  & 0.29& 0.22 & 0.14\\
\cline{1-8}
3 $M_\odot$ [Fe/H]$=-0.3$&ST &1.07 &97  &0.47  &1.14&1.15&1.10\\
&ST/1.5 &1.07& 97 & 0.46 & 0.96 & 0.89 & 0.76 \\
&ST/3& 1.07 & 97 & 0.22 & 0.48 & 0.40& 0.32\\ 
\cline{1-8}
5 $M_\odot$ [Fe/H]$=+0.0$&ST &1.00 &92  &1.33  &1.11&1.08&1.29\\
&ST/1.5 &1.00& 92 & 1.24 & 0.93 & 0.87 & 1.02 \\
&ST/3& 1.00 & 92 & 1.04 & 0.71 &  0.63& 0.73 \\ 
\cline{1-8}
5 $M_\odot$ [Fe/H]$=-0.3$&ST& 1.01 &93  &0.92  &0.91&0.96&1.29\\
&ST/1.5 &1.01& 93 & 1.02 & 0.89 & 0.89 & 1.12 \\
&ST/3& 1.01 & 93 & 0.96 & 0.67 &  0.61& 0.74 \\
\enddata
\end{deluxetable}
\clearpage

\figcaption[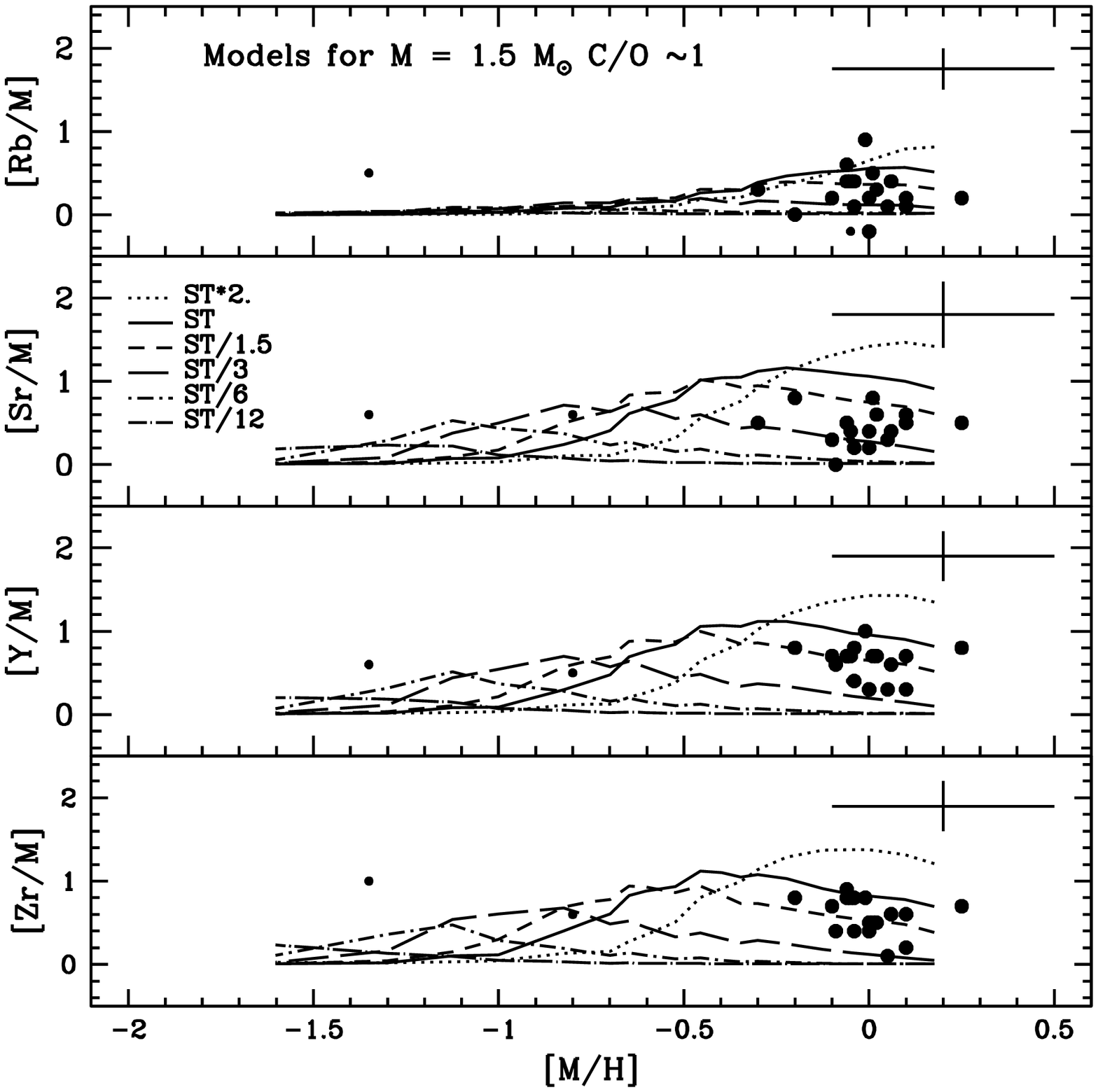]{Derived abundance ratios [X/M] vs. [M/H] in our sample
of C-stars (dots) compared with the corresponding theoretical predictions (lines)
for a 1.5 \msb TP-AGB star when the envelope reaches C/O $\ga$ 1 for different
parameterizations of the $^{13}$C pocket (see text). A typical error bar is shown. 
In the stars where a very few number of lines are used the
error bar is certainly larger. Note, furthermore, that the data points
for IY Hya, VX Gem and V CrB are plotted with smaller symbols to indicate that 
the abundance ratios derived in them are very uncertain.\label{fig1}} 

\figcaption[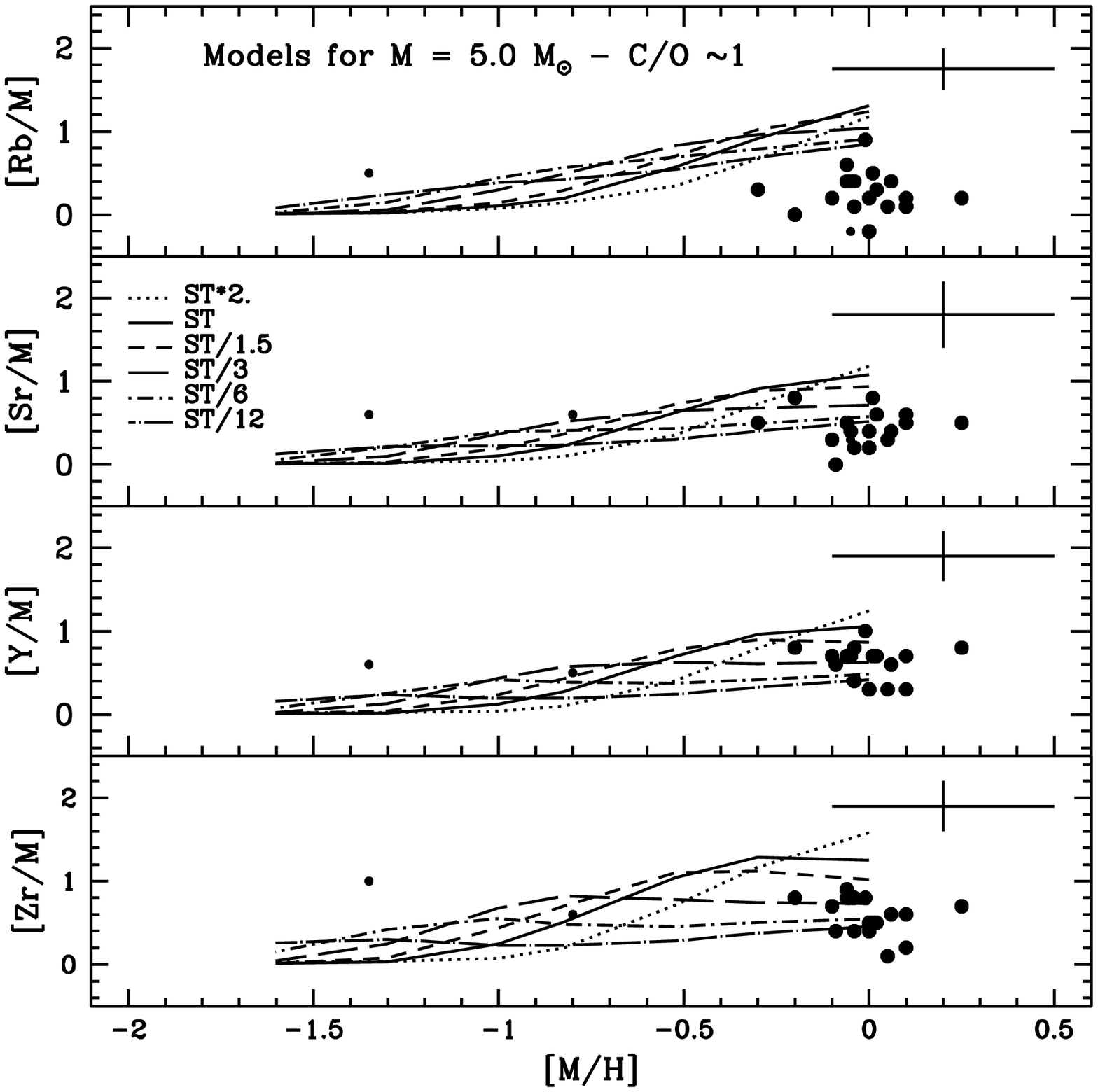]{As Figure 1 comparing with theoretical predictions
for a 5 \msb TP-AGB star.\label{fig2}}

\figcaption[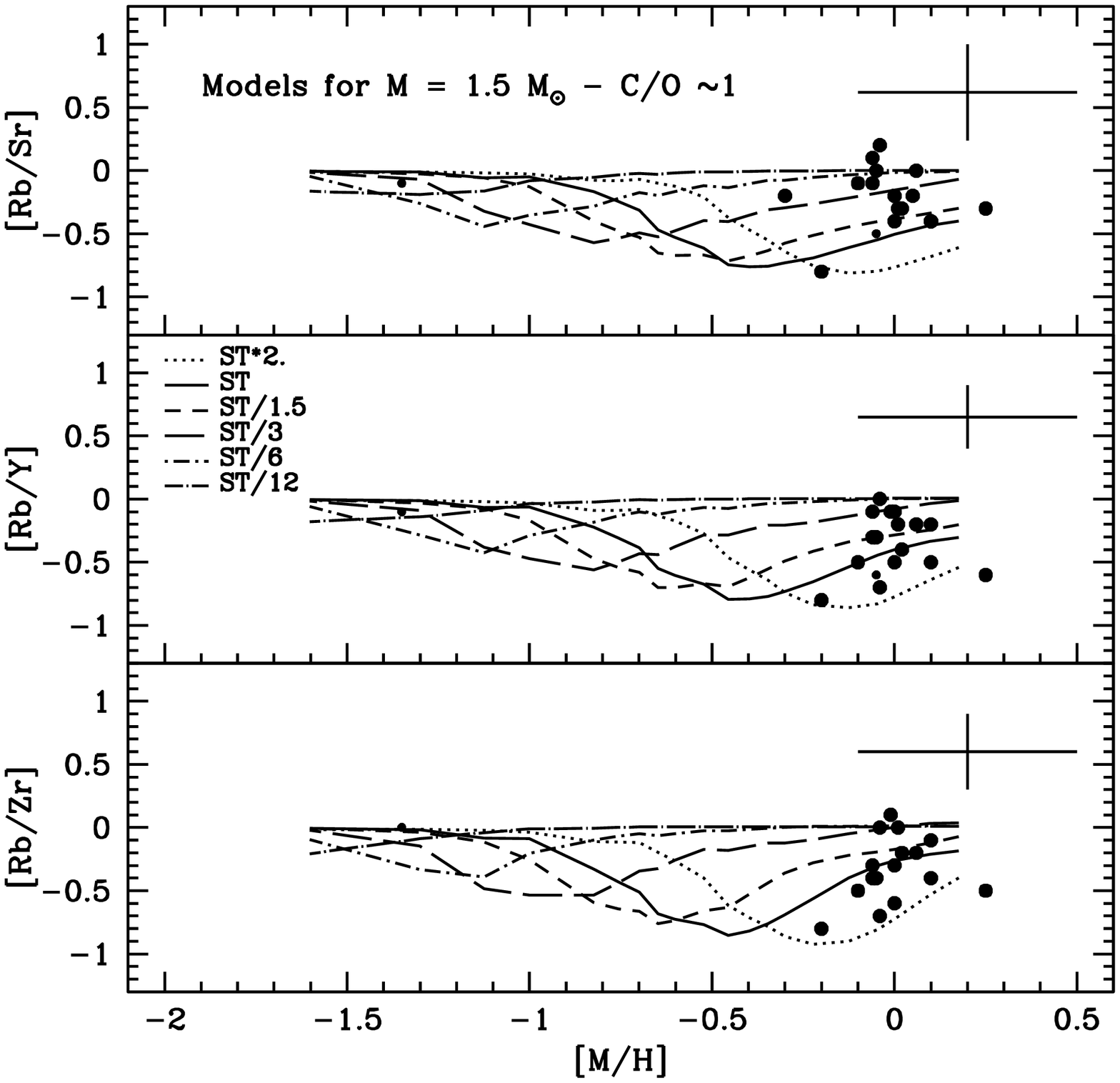]{Comparison of the observed [Rb/Sr,Y,Zr] ratios
vs. [M/H] with different theoretical predictions for a 1.5 \msb TP-AGB star
when the envelope reaches C/O $\ga$ 1. Note that again the data points for
IY Hya, VX Gem and V CrB are plotted with smaller symbols.\label{fig3}}

\figcaption[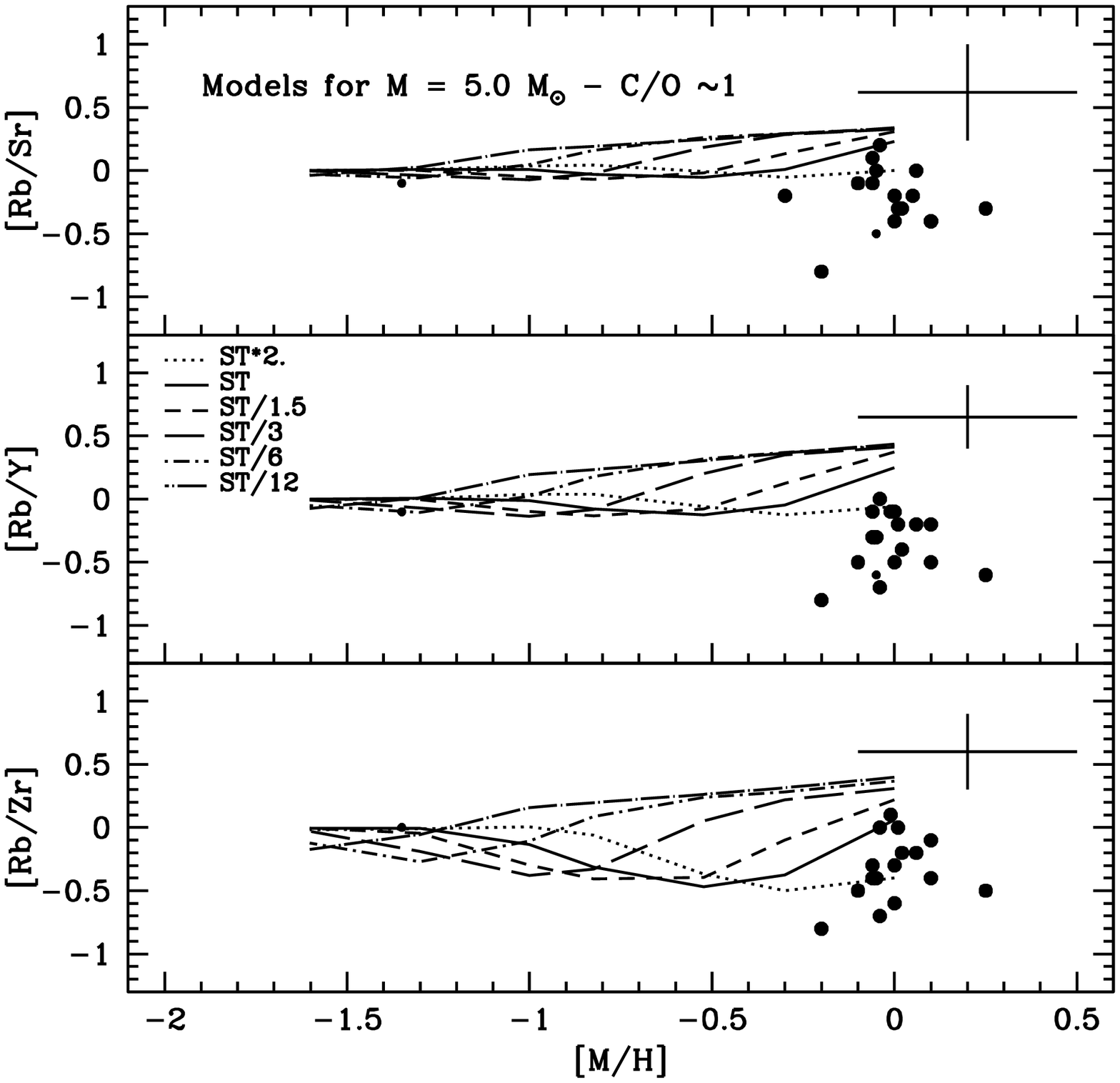]{As Figure 3 comparing with theoretical predictions
for a 5 \msb TP-AGB star.\label{fig4}}

\figcaption[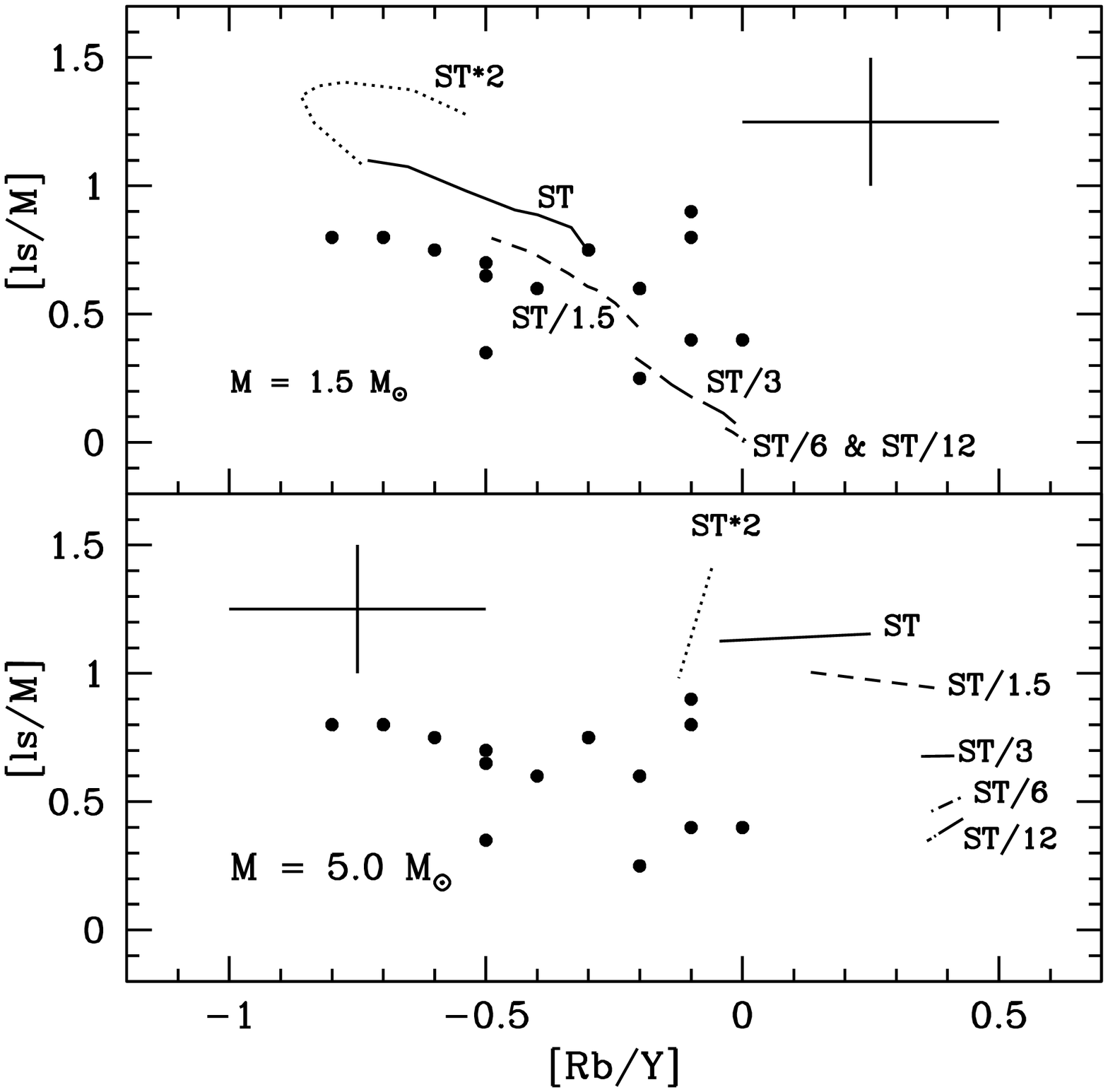]{Comparison of the observed mean low-mass s-element
(Y,Zr) enhancement (signature of the neutron exposure) against [Rb/Y] 
(signature of the neutron density) with theoretical predictions for a
1.5 \msb (upper panel) and 5 \msb (lower panel) TP-AGB star for different
parameterizations of the $^{13}$C pocket (lines). The theoretical predictions
shown are for stellar models with C/O $\sim$ 1 and 
metallicity [Fe/H] $\ga$ $-0.3$, 
which is the metallicity of most of the stars studied here. Only stars in the sample 
with metallicity above this value are plotted (see text). Note that several 
stars coincide in the same data point.\label{fig5}}
 
\figcaption[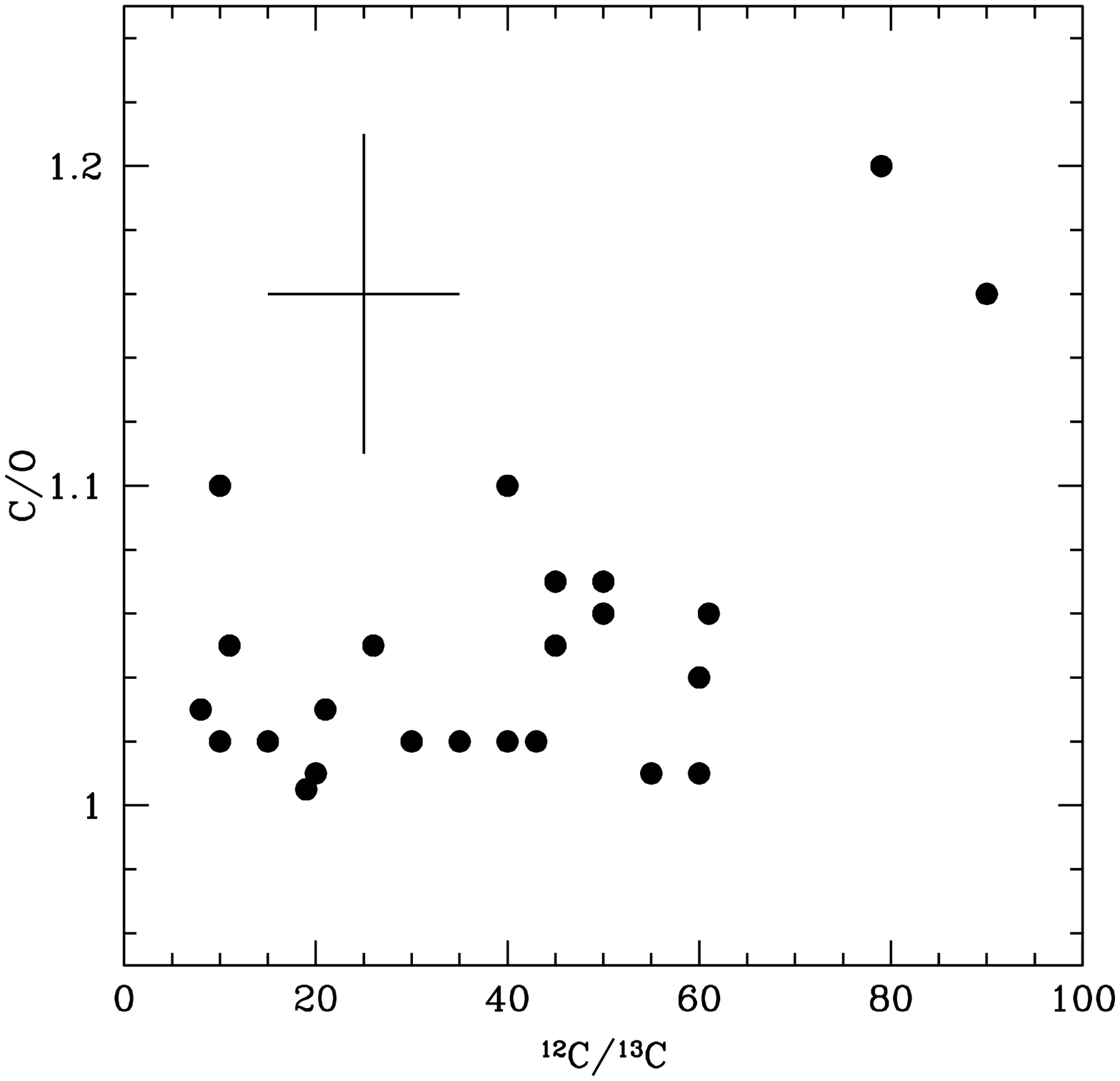]{C/O vs. $^{12}$C/$^{13}$C ratios derived
in our stars. Note that most of the stars have a C/O ratio only slightly higher than
1 and that a significant number show $^{12}$C/$^{13}$C $<$ 40.   
An average error bar is shown.\label{fig6}} 
 
\clearpage
\plotone{f1.eps}
\clearpage
\plotone{f2.eps}
\clearpage
\plotone{f3.eps}
\clearpage
\plotone{f4.eps}
\clearpage
\plotone{f5.eps}
\clearpage
\plotone{f6.eps} 

\end{document}